\documentclass[10pt,twocolumn,letterpaper]{article}

\usepackage{iccv}
\usepackage{times}
\usepackage{epsfig}
\usepackage{graphicx}
\usepackage{amssymb}

\usepackage{amsmath,amsthm}
\usepackage{algorithm}
\usepackage{algpseudocode}

\usepackage{graphicx}
\usepackage{caption}
\usepackage[accsupp]{axessibility} 

\input{shortcut.tex}

\newcommand\yiling[1]{\textcolor{black}{#1}}
\newcommand\agao[1]{\textcolor{black}{#1}}
\usepackage[pagebackref=true,breaklinks=true,letterpaper=true,colorlinks,bookmarks=false]{hyperref}

\iccvfinalcopy % *** Uncomment this line for the final submission

\def\iccvPaperID{2443} % *** Enter the ICCV Paper ID here

\ificcvfinal\fi

\begin{document}

%%%%%%%%% TITLE
\title{Dynamic Mesh-Aware Radiance Fields}

\author{Yi-Ling Qiao$^*$~~~~Alexander Gao$^*$~~~~Yiran Xu~~~~Yue Feng~~~~Jia-Bin Huang~~~~Ming C. Lin\\
% University of Maryland, College Park\\
\vspace{2mm}
{University of Maryland College Park} \\
% \jiabin{Please add a project website here, for example: https://mesh-aware-rf.github.io}
\url{https://mesh-aware-rf.github.io}
% \vspace{2mm}
% {\texttt{yilingq, awgao, yiranx, yuefeng, jbhuang, lin\}@cs.umd.edu}
% \and
}

\newcommand\blfootnote[1]{%
  \begingroup
  \renewcommand\thefootnote{}\footnote{#1}%
  \addtocounter{footnote}{-1}%
  \endgroup
}

%     \includegraphics[width=\linewidth]{figure/teaser.pdf}
%     \vspace{-2mm}
% \captionof{figure}{ \textbf{Mesh-aware rendering of radiance fields.}
% In the scene \emph{Garden}~\cite{barron2022mipnerf360} modeled as a radiance field, we place a cubic mesh with reflective textures and other synthetic mesh objects. 
% Our mesh-aware rendering explicitly computes the rays bouncing inside the mirror room, creating an `infinite mirror room' visual effect.
% }
%     \label{fig:teaser}

%     \includegraphics[width=\linewidth]{figure/teaser2.pdf}
%     \vspace{-2mm}
% \captionof{figure}{ \textbf{Mesh-aware rendering of radiance fields.}
% We place a cubic mesh with reflective textures and other synthetic mesh objects in the scene \emph{Garden}~\cite{barron2022mipnerf360} modeled as a radiance field. 
% Our mesh-aware rendering explicitly computes the rays bouncing inside the mirror room, creating an `infinite mirror room' visual effect.
% }
%     \label{fig:teaser}

\twocolumn[{%
\renewcommand\twocolumn[1][]{#1}%
\maketitle
\begin{center}
    \centering
    \captionsetup{type=figure}
    \includegraphics[width=\linewidth]{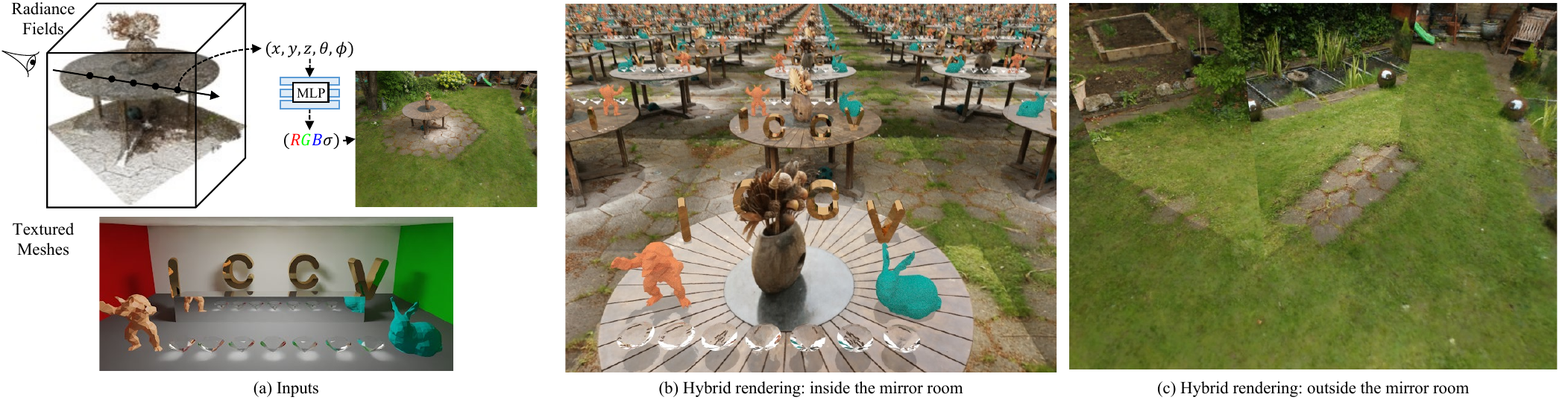}
    \captionof{figure}{ \textbf{Mesh-aware rendering of radiance fields.}
We place a cubic mesh with reflective textures and other synthetic mesh objects in the \emph{MipNeRF-360 Garden}~\cite{barron2022mipnerf360} scene. Our mesh-aware rendering explicitly computes the rays bouncing inside the mirror room, creating an `infinite mirror room' visual effect.
}
    \label{fig:teaser}
\end{center}%
}]

% \twocolumn[{%
% \renewcommand\twocolumn[1][]{#1}%
% \maketitle
% \begin{center}
%     \centering
%     \captionsetup{type=figure}
%     \includegraphics[width=\linewidth]{figure/teaser.pdf}
%     \captionof{figure}{\textbf{Mesh-aware rendering of radiance fields.}
% In the scene \emph{Garden}~\cite{barron2022mipnerf360} modeled as a radiance field, we place a cubic mesh with reflective textures and other synthetic mesh objects. 
% Our mesh-aware rendering explicitly computes the rays bouncing inside the mirror room, creating an `infinite mirror room' visual effect.}
%     \label{fig:teaser}
% \end{center}%
% }]

% \maketitle
\ificcvfinal\thispagestyle{empty}\fi

%%%%%%%%% ABSTRACT
\begin{abstract}
% Reconstructing and setting up the background scenes for rendering used to be costly and require artistic skills. 
Embedding polygonal mesh assets within photorealistic Neural Radience Fields (NeRF) volumes, such that they can be rendered and their dynamics simulated in a physically consistent manner with the NeRF, is under-explored from the system perspective of integrating NeRF into the traditional graphics pipeline. This paper designs a two-way coupling between mesh and NeRF during rendering and simulation. We first review the light transport equations for both mesh and NeRF, then distill them into an efficient algorithm for updating radiance and throughput along a cast ray with an arbitrary number of bounces. To resolve the discrepancy between the linear color space that the path tracer assumes and the sRGB color space that standard NeRF uses, we train NeRF with High Dynamic Range (HDR) images. We also present a strategy to estimate light sources and cast shadows on the NeRF.  Finally, we consider how the hybrid surface-volumetric formulation can be efficiently integrated with a high-performance physics simulator that supports cloth, rigid and soft bodies. The full rendering and simulation system can be run on a GPU at interactive rates.  We show that a hybrid system approach outperforms alternatives in visual realism for mesh insertion, because it allows realistic light transport from volumetric NeRF media onto surfaces, which affects the appearance of reflective/refractive surfaces and illumination of diffuse surfaces informed by the dynamic scene.
\blfootnote{*Equal contribution}
% users can build their own virtual world simply with there phone camera and 
\end{abstract}

%%%%%%%%% BODY TEXT
\section{Introduction}
% \todo{[Experiments]: comparison rendering x 3, simulation x 2} \\
% \todo{[Experiments]: Teaser} \\
% % \todo{[Writing]} \\
% \yiling{[Writing] Experiments} \\

% \begin{figure}[t]
%     \centering
%     \includegraphics[width=1\linewidth]{figure/teaser_dark.jpg}
%     \caption{Mirros.}
%     \label{fig:teaser}
% \end{figure}

% Building a photorealistic virtual world is a core task of computer graphics. 
Creating high-quality 3D environments suitable for photorealistic rendering entails labor-intensive manual work carried out by skilled 3D artists.
Neural Radiance Fields (NeRF)~\cite{mildenhall2020nerf} provide a convenient way to capture a volumetric representation of a complex, real-world scene, paving the way for high-quality novel view synthesis and interactive photorealistic rendering~\cite{mueller2022instant}. 
These qualities make NeRF exceptionally adept at modeling background environments.
On the other hand, existing methods for physically-based simulation and rendering of complex material and lighting effects are primarily based on \emph{surface mesh} representations.  
%How to best incorporate NeRF's volumetric scene representations with widely adopted traditional graphics pipelines remains unclear.
\yiling{Integrating neural field representations with well-established traditional graphics pipelines opens up many possibilities in VR/AR, interactive gaming, virtual tourism, education, training, and computer animation.}

% Challenges of embedding NeRF to simulation/rendering pipelines
% a. Rendering: continuous (NeRF) v.s. discrete (path-tracing)
% b. NeRF density field can't give good surfaces
% c. Vanilla NeRF is a slow renderer.
\yiling{Volume rendering~\cite{novak14thesis} has demonstrated its capability to produce visually captivating results for participating media~\cite{muller16efficient}. 
However, integrating NN-based NeRF into this pipeline while maintaining realistic lighting effects such as shadows, reflections, refractions, and more, remains a relatively unexplored area.}
% Attempting to unify implicit, continuous NeRF with explicit, discrete mesh representations in an integrated simulation and rendering pipeline presents new challenges. The ray-marching rendering of NeRF is integrated over a continuous ray, while path tracing over meshes is composed of discrete ray bounces. 
In terms of simulation, while the geometry of NeRF is implicit in its density field, it lacks a well-defined surface representation, making it difficult to detect and resolve collisions.
Recent works have delved into enhancing the integration between NeRF and meshes, aiming to combine the photorealistic capabilities of NeRF with the versatility of meshes for rendering and simulation. Neural implicit surfaces~\cite{yariv2021volume,wang2021neus,Oechsle2021unisurf,esposito2022kiloneus} are represented as learned Signed Distance Fields (SDF) within the NeRF framework. Meanwhile, methods like IRON~\cite{zhang2022iron} and NVDiffRec~\cite{munkberg2021nvdiffrec} extract explicit, textured meshes that are directly compatible with path tracing, offering practical benefits at the expense of a lossy discretization. 
%\yiran{kinda ambiguous to me, perhaps say ``do not \emph{directly} render meshes with NeRF''?} 
Nerfstudio~\cite{nerfstudio} renders NeRF and meshes separately, then composites the render passes with an occlusion mask. Unfortunately, this decoupled rendering approach offers no way to exploit the lighting and appearance information encoded in the NeRF volume to affect the rendered mesh appearance. Figure~\ref{fig:comp_render} visually compares our hybrid method to naively combining NeRF and surface rendering, and pure surface rendering.

% which we will show in our example experiments.

% \begin{figure}[t]
% \centering
% \begin{tabular}{@{}c@{\hspace{1mm}}c@{}}
%     % \includegraphics[width=0.32\linewidth]{figure/mirror_ours.jpg}&
%     \includegraphics[width=0.5\linewidth]{figure/mirror_empty.jpg}   &
%     \includegraphics[width=0.5\linewidth]{figure/mirror_ours.jpg} 
%     \\
%   \small   \small (a) Original NeRF~\cite{nerfstudio} & \small (b) Ours \\
%     \includegraphics[width=0.5\linewidth]{figure/nerfstudio_infinity_sigasia.png}   &
%     \includegraphics[width=0.5\linewidth]{figure/mirror_nvdifrecc_monkey.png} 
%     \\
%   \small    (c) Nerfstudio~\cite{nerfstudio} & \small (d) NvDiffRec~\cite{munkberg2021nvdiffrec} \\ 
% \end{tabular}
% % \vspace{-1.0em}
% \caption{
% \yiling{
% \textbf{Limitations of alternative methods.} 
% Rendering the NeRF in an infinite mirror room.
% % (a) Our method can correctly simulate the radiance field and mesh together. 
% (a) Nerfstudio~\cite{nerfstudio} renders the radiance field and mesh separately, then composites them with visibility maps.  
% (b) NvDiffRec~\cite{munkberg2021nvdiffrec} reconstructs a mesh from NeRF and renders everything in Blender, but reconstruction usually has lower quality than the original NeRF. }
% }
% % \vspace{-1.5em}
% \label{fig:comp_mirror}
% \end{figure}
\begin{figure}[t]
    \centering
    % \vspace{-1em}
    \includegraphics[width=1\linewidth]{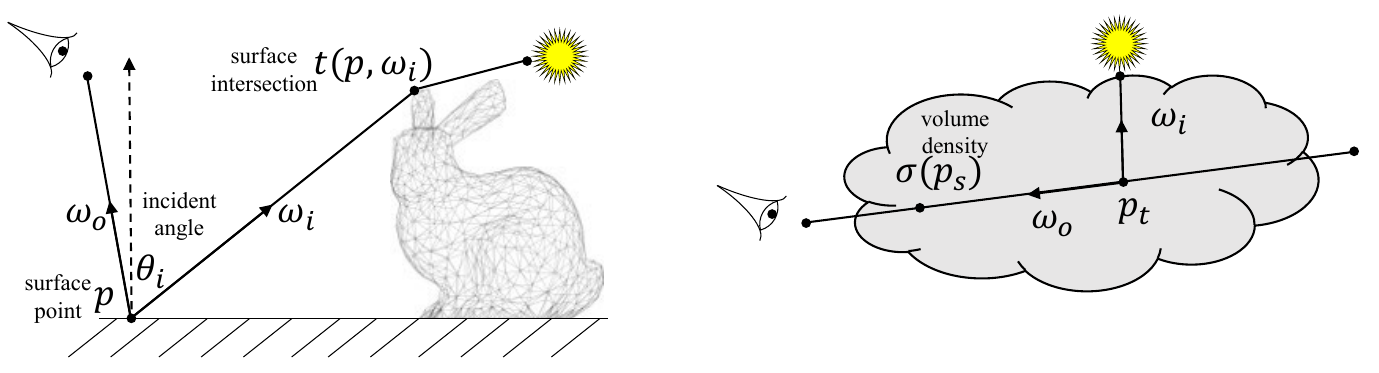}
\caption{\textbf{Light transport} on the surface (\emph{left}) and in the medium (\emph{right}).}
\label{fig:light_transport}
\vspace{-1em}
\end{figure}

% Our contributions
We introduce a hybrid graphics pipeline that integrates the rendering and simulation of neural fields and meshes. for both representations, we consider lighting effects and contact handling for physical interaction. \yiling{By unifying NeRF volume rendering and path tracing within the linear RGB space, we discover their Light Transport Equations exhibit similarities in terms of variables, forms, and principles. Leveraging their shared light transport behavior, we devise update rules for radiance and throughput variables, enabling seamless integration between NeRF and meshes.}
% \yiran{It is unclear why we need to examine similar variables from the context. What's the connection between the similarities and the accumulation below? }
\yiling{To incorporate shadows onto the NeRF, we employ differentiable surface rendering techniques~\cite{Mitsuba3} to estimate light sources and introduce secondary shadow rays during the ray marching process to determine visibility. Consequently, the NeRF rendering equation is modified to include a point-wise shadow mask.}

For simulation, we adopt SDFs to represent geometry of neural fields, which is advantageous for physical contact handling and collision resolution. We then use position-based dynamics~\cite{Macklin2022XPBD} for time integration. Our efficient hybrid rendering and simulation system is implemented in CUDA. To enhance usability, we have also incorporated user-friendly Python interfaces. In summary, the key contributions of this work are:
%\vspace{-\topsep}
\begin{itemize}
\setlength{\itemsep}{1mm}
\setlength{\parskip}{1pt}
    \item A two-way coupling between NeRF and surface representations for rendering and simulation.
    \item Integration with HDR data which can unify the color space of the path tracer and NeRF, with a strategy to estimate light sources and cast shadows on NeRF.
    \item An efficient rendering procedure that alternates ray marching and path tracing steps by blending the Light Transport Equations for both NeRF and meshes.
    % \item Use of Signed Distance Field (SDF) as NeRF's representation for its compatibility with neural fields and ability to model complex topology; % The scene is simulated by position-based dynamics with hybrid SDF and mesh objects. \agao{could mention that SDF works well for this purpose, but other approaches could also work. most importantly, we show how to integrate the reconstruction/simulation into the hybrid rendering pipeline (NOT claiming to advance SOTA of reconstruction in this paper)}
    \item An interactive, easy-to-use implementation with a high-level Python interface that connects the low-level rendering and simulation GPU kernels.
\end{itemize}

\section{Related Work}
% Corrected
% Saved
% 430 words
\subsection{Neural Fields and Surface Representations}
\begin{figure*}[t]
    \centering
    % \vspace*{-0.5em}
    \includegraphics[width=1\linewidth]{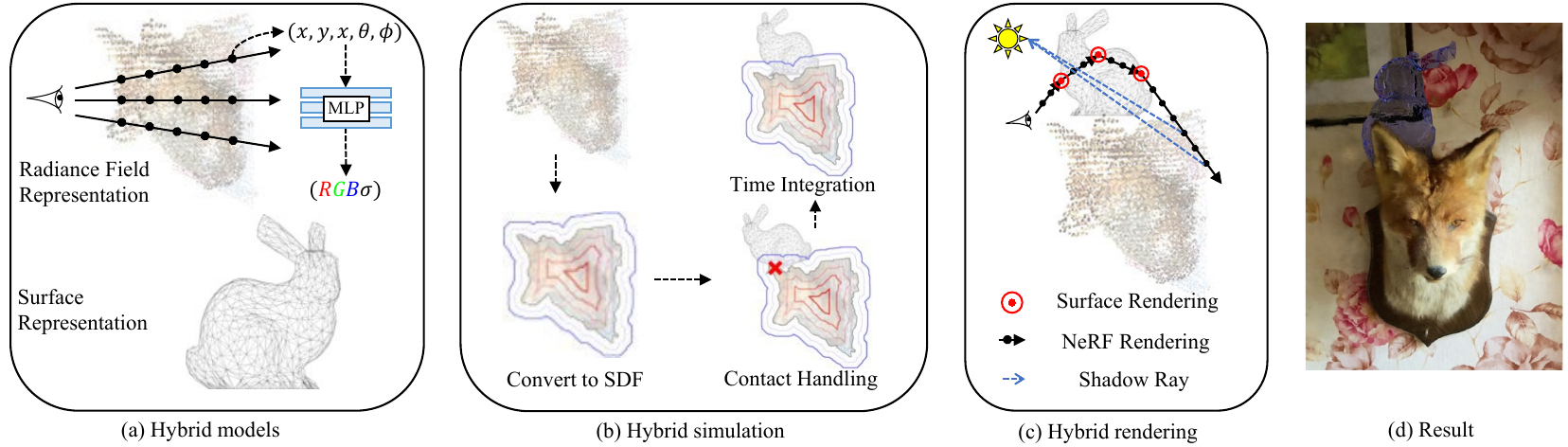}
% \vspace{-1em}
\caption{ 
\textbf{Pipeline overview.} 
Our method takes an optimized radiance field model and surface meshes as inputs. We can run a physics simulation between the NeRF and meshes. The updated mesh vertices and NeRF transformations are synchronized to the renderer, which uses Monte Carlo simulation to sample ray paths. As the ray travels through space, it alternates between surface rendering (ray-tracing) and NeRF rendering (ray-marching), both updating its radiance.
}
    \label{fig:overview}
    % \vspace*{-0.5em}
\end{figure*}

\yiling{Rendering of participating media has been extensively studied ~\cite{Novak18volumeSTAR,muller16efficient} in classic graphics pipelines~\cite{Novak18Course,pharr2016physically,Mueller2016HGM,hua2019survey}. 
In recent years, significant advancements have been made in this area~\cite{miller19null,nimierdavid2022unbiased}, yielding remarkable visual outcomes. Our work aims to expand upon this progress by incorporating the emergent Neural Radiance Fields (NeRF)~\cite{mildenhall2020nerf}, which have gained substantial popularity, into this exciting domain.}
Within the volume rendering framework, NeRF bakes the plenoptic function and volumetric density into spatial points. These points can be effectively parameterized by an an MLP~\cite{mildenhall2020nerf}, convolutional networks~\cite{Chan2021}, hash grid~\cite{mueller2022instant}, point cloud~\cite{xu2022point}, voxel~\cite{liu2020neural,SunSC22}, or tensors~\cite{Chen2022ECCV}. It allows users to reconstruct photorealistic 3D static~\cite{tancik2022block,turki2022mega,rematas2022urban,kopanas2022neural,wang2021mirrornerf} or dynamic scenes~\cite{park2021hypernerf,xian2021space,park2021nerfies,gao2021dynamic,liu2023robust} by casually capturing a few images~\cite{Luma2022}. 
Original NeRF takes seconds to render one single frame, while follow-up works have accelerated rendering speed~\cite{hedman2021baking,wang2022r2l,chen2022mobilenerf,garbin2021fastnerf,mueller2021,yu2021plenoxels,yu2021plenoctrees,attal2022learning,attal2023hyperreel,wang2022fourier}.

NeRF simplifies the creation of 3D content compared to classical mesh-based pipelines. However, addressing challenges of editing~\cite{zhang2021editable,kasten2021layered} and decomposing the baked information~\cite{martin2021nerf,rudnev2021neural,srinivasan2021nerv,yang2021learning} is not trivial. Effort has been directed toward reconciling the advantages of both NeRF and surface-based paradigms. In rendering, \cite{zhang2022iron} and \cite{munkberg2021nvdiffrec} propose to use surface-based differentiable rendering to reconstruct textured meshes~\cite{hasselgren2022nvdiffrecmc} from neural fields. Their reconstructed meshes can be imported to a surface rendering pipeline like Blender~\cite{blender}, but the original NeRF representation cannot be directly rendered with meshes. 
For simulation, NeRFEditting \cite{Yuan22NeRFEditing} proposes to use explicit mesh extracted by~\cite{wang2021neus} to control the deformation of Neural Fields. Qiao et al. \cite{qiao2022neuphysics} further add full dynamics over the extracted tetrahedra mesh. Chu et al. \cite{chu2022physics} integrates the dynamics of smoke with neural fields. \cite{Cleac2022Differentiable} also connects differentiable simulation to NeRF, where the density field and its gradient are used to compute the contact. These methods aim to construct an end-to-end differentiable simulation and rendering pipeline, yet they have yet to couple the rendering.

\subsection{Scene Editing and Synthesis}
Our method enables inserting mesh assets into NeRF models of real-world captures. Editing of existing scenes is an active topic of study. For neural field representations, ray bending~\cite{tretschk2021nonrigid,pumarola2020dnerf,li2023pac} is widely used to modify an optimized NeRF. It is possible to delete, add, duplicate, or actuate~\cite{chen2021animatable,peng2021neural,weng_humannerf_2022_cvpr,liu2021neural} an area by bending the path of the rendering ray. \cite{guo2022objectcentric} propose to train a NeRF for each object and compose them into a scene. ClimateNeRF~\cite{li2022climatenerf} can change weather effects by modifying the density and radiance functions during ray marching. These methods study editing of isolated NeRF models. There are also inverse rendering works that decompose~\cite{yao2022neilf,boss2021nerd} the information baked into NeRF, which can then be used to edit lighting and materials~\cite{kobayashi2022distilledfeaturefields}. 
Such decomposition is useful, but assumes information like a priori knowledge of light sources, or synthetic scenes. They do not address inserting mesh into NeRF scenes. Besides NeRF, \cite{Karsch2011Rendering} inserts a virtual object into existing images by estimating the geometry and light source in the existing image. \cite{chen2021geosim} insert vehicles in street scenes by warping textured cars using predicted 3D poses.

% \subsection{Simulation}

% rigid body

% cloth

% deformable solids

% 12
% All suggestions

\section{Method}
In this section, we describe how radiance fields and polygonal meshes can be integrated into a unified rendering and simulation pipeline, an overview of which is visualized in Figure~\ref{fig:overview}.
\begin{figure*}
\centering
\includegraphics[width=\linewidth]{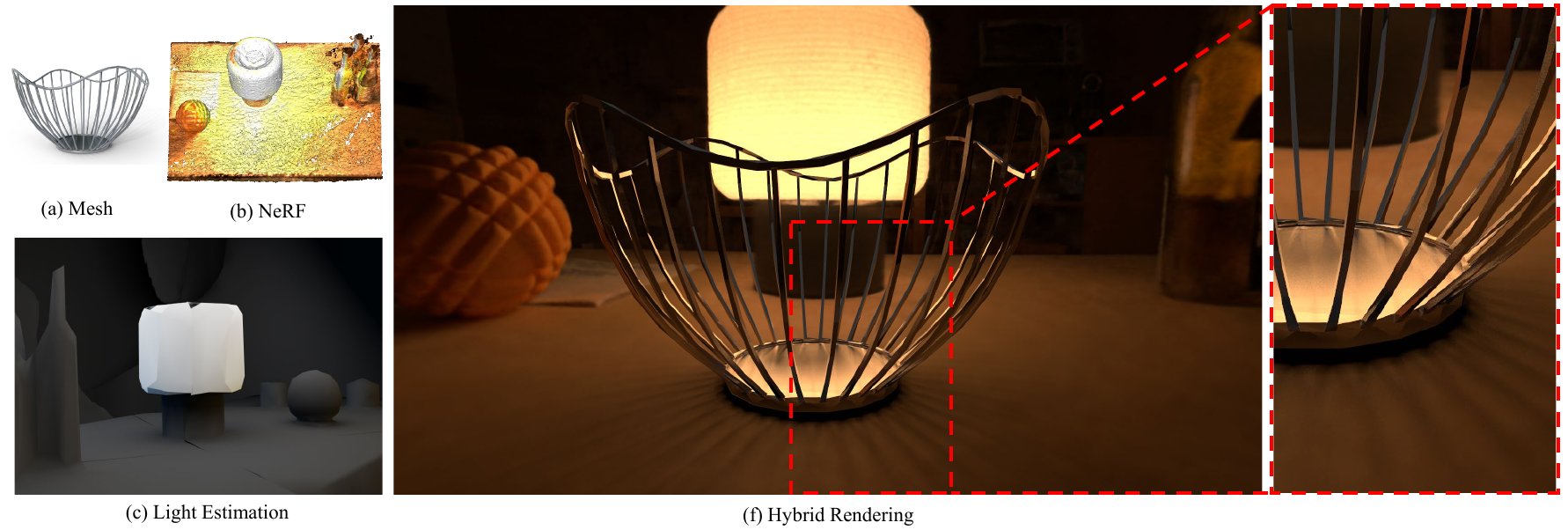}
% \begin{tabular}{@{}c@{\hspace{1mm}}c@{}}
%     \includegraphics[width=0.5\linewidth]{example-image} &
%     \includegraphics[width=0.5\linewidth]{example-image} \\
%     \includegraphics[width=0.5\linewidth]{example-image} &
%     \includegraphics[width=0.5\linewidth]{example-image} \\
%     \small (c) Orignal Scene & \small (d) Inserted mesh with shadow 
% \end{tabular}
% \vspace{-1em}
\caption{\textbf{Shadow Casting.} We estimate the geometry and light source of the scene and insert a metal basket onto the desk. Our pipeline can render the realistic reflection and shadow effects caused by the synthetic mesh.}
\label{fig:shadow_ablation}
% \vspace{-1em}
\end{figure*}

\subsection{Rendering}
NeRF can photorealistically reconstruct a 3D scene from a set of images, making it an appealing candidate for modeling environments and potentially valuable for traditional surface-based rendering.  One possible approach to bridge these disparate representations is to render the meshes and NeRF volume in separate passes, and composite the results together in 2D image space. However, compositing in image space is susceptible to incorrect occlusion masks and inaccurate lighting. 
A more physically principled approach to this problem is identifying and exploiting the similarities in their respective light transport equations, which directly allows the radiance field and mesh to be incorporated in 3D space. \\

% \agao{Before launching into the equations, restate the main goals of the hybrid rendering system to support why it is useful to think about the problem and solution in this way, as opposed to other approaches (e.g. why do we need to design a hybrid renderer instead of just using separate mesh and volume renderers and trying to composit the separate results together?).  \textit{Mention: finding the similarity between their light transport equations allows us to incorporate them in a way that is based on physical principles of light}}

\mypara{Surface Rendering Equation.}
The Light Transport Equation (LTE) for surface rendering is:
\begin{equation}
\label{eq:SLTE}
    L(p,\omega_o) = L_e(p,\omega_o) + L_r(p, \omega_o)
\end{equation}

\vspace{-1em}
\begin{equation}
\label{eq:reflected}
    L_r(p, \omega_o) = \int_{S^2}f_s(p,\omega_o, \omega_i)L_i(p, \omega_i)\abs{\cos\theta_i}d\omega_i
\end{equation}
% \vspace{+0.5em}

\begin{equation}
\label{eq:cast}
    L_i(p, \omega_i) = L(t(p, \omega_i), -\omega_i)
\end{equation}
% \vspace{+0.5em}
where $p$ is a surface point; $\omega_i,\omega_o$ are the directions of incident (incoming) and exitant (outgoing) radiance; $S^2$ is the unit sphere sampling space for directions; $L , L_e, L_i, L_r$ are the exitant, emitted, incident, and reflected radiance, respectively; $\theta_i$ is the angle of incidence of illumination; $f_s(p,\omega_o, \omega_i)$ is the bidirectional scattering distribution function (BSDF); and $t(p, \omega)$ is the ray-casting function that computes the first surface intersected by the ray cast from $p$ in the direction $\omega$. %\yiling{do we also need to define $f_s$, $\theta$, and $S^2$?}

If a scene is represented solely by surfaces, the LTE in Equation~\ref{eq:SLTE} can be solved by Monte Carlo path tracing: for each pixel, a ray is randomly cast from the camera, its path constructed incrementally each time it hits, and bounces off of a surface. A natural but memory-inefficient way to implement this algorithm is to recursively compute Equation~\ref{eq:SLTE} and spawn a new ray upon each ray-surface intersection. Noticing that $L_i(p,\omega_i)$ is independent of previous paths, the recursive process can be transformed into a weighted sum of radiance on each ray-surface intersection $p_k$. These weights $T(p_k)$ are called \textit{throughput}, and they depend on their predecessors' BSDF $f_s(p,\omega_o, \omega_i)$, illumination angle $\abs{\cos\theta_i}$, and probability density function $P$ of the scattering: 
\begin{equation}
\label{eq:path_t}
    T(p_{k}) = \frac{T(p_{k-1})\cdot f_s(p_{k}, \omega_{k-1}, \omega_{k})\abs{\cos(\theta_k)}}{P}
\end{equation}
% More details of the throughput function can be found in Appendix (xxx).

$T(p_k)$ and $L_i(p_k,\omega)$ are the only variables essential to track and integrate for each bounce. \\

\mypara{Volumetric Rendering Equation.}
The light transport equation for the volumetric medium is:
% emission
\begin{equation}
\label{eq:VLTE}
    L(p,\omega_o) = \int_{t=0}^{t_f}\exp\Big(-\int_{s=0}^t\sigma_t(p_s)ds\Big)L_i(p_t,\omega_o)dt
\end{equation}
% \vspace{-0.5em}
\begin{equation}
\label{eq:VLTE_Li}
\begin{aligned}
    L_i(p_t,\omega_o) = & L_e(p_t, \omega_o) + L_s(p_t, \omega_o)\\
\end{aligned}
\end{equation}
\vspace{-2em}

\begin{equation}
\label{eq:VLTE_Li}
% \vspace{-0.5em}
\begin{aligned}
    L_s(p_t, \omega_o) = \sigma_s(p_t)\int_{S^2}f_p(p_t,\omega_o,\omega_i)L(p_t, \omega_i)d\omega_i
\end{aligned}
\end{equation}
where $L_s$ is the (weighted) scattered radiance, $\sigma_t$ and $\sigma_s$ are the attenuations and scattering coefficients, $f_p$ is the phase function of the volume, and $p_s$ is on the ray $p_s=p+s\cdot\omega_o$ (similar to $p_t$).  All other terms share the same definition as in surface rendering.

The integral in the volumetric LTE could again be solved using Monte Carlo methods. However, stochastic simulation of volumetric data is more challenging and expensive than surface data. A photon may change direction in a continuous medium, unlike the discrete bounces that occur only at surfaces. Therefore, rather than simulating the path of photons using Monte Carlo sampling, methods like NeRF~\cite{mildenhall2020nerf} instead bake the attenuation coefficient $\sigma(p)=\sigma_t(p)$ and view-dependent radiance $r(p,\omega)$ onto each spatial point, and so there is no scattering.  This circumvents solving Equation~\ref{eq:VLTE_Li}, thereby avoiding considering light transport, light sources, and material properties.
%\yiling{can we make more comparison between the volumetric LTE and the NeRF rendering equation? i.e. why do they use different terms like $L$ vs. $r$, and are they interchangeable?} 
Volume rendering under the NeRF formulation becomes:
\begin{align}
\label{eq:nerf_rendering}
    r(p, \omega) &= \int_{0}^{t_f} T(t)\sigma(p_t) r(p_t, \omega) dt, \\
    T(t)&=\exp\Big(-\int_{0}^t\sigma(p_s)ds\Big)
\end{align}
% While baking in the attenuation and view dependence makes the volume rendering more efficient, it also loses some of the raw information of the scene. In order to reconstruct the scene for relighting and editing, one must perform intrinsic decomposition on NeRF~\cite{srinivasan2021nerv}. Solving the inverse rendering problem is not within the scope of this work. Rather, we focus on providing a way to incorporate existing vanilla NeRF (e.g. a pre-trained Instant-NGP) into surface-based rendering and simulation. More complicated features like shadows and global illumination are left for future work.
In Equation ~\ref{eq:nerf_rendering}, the radiance and throughput are being updated, similar to surface rendering.  However, note that the volumetric LTE denotes incident radiance at point $p$ from direction $\omega$ as $L_i(p,\omega)$, while NeRF denotes the same as $r(p,\omega)$.  This terminology is indeed overloaded, as $r(p,\omega)$ in the NeRF formula represents sRGB color, i.e. the result of applying a nonlinear tone-mapping function to the raw radiance value.  The terms are related in that $r(p, \omega) = \psi(L_i(p,\omega))$, where $\psi(\cdot)$ represents a tone-mapping function from linear to sRGB color space.

% \begin{figure}
% \centering
% \begin{tabular}{@{}c@{\hspace{1mm}}c@{}}
%     \includegraphics[width=0.5\linewidth]{example-image} &
%     \includegraphics[width=0.5\linewidth]{example-image} \\
%     \includegraphics[width=0.5\linewidth]{example-image} &
%     \includegraphics[width=0.5\linewidth]{example-image} \\
%     \small (c) Orignal Scene & \small (d) Inserted mesh with shadow 
% \end{tabular}
% \vspace{-1em}
% \caption{\textbf{Inverse rendering for shadow casting.} }
% \label{fig:inverse_rendering}
% \vspace{-1em}
% \end{figure}
\mypara{Unifying color space of path tracing and ray marching.} The standard NeRF model accumulates sRGB color along rays cast from the camera into the volume: each point's color is represented by three 8-bit values, one for each color channel.  Integrating these colors along the ray (weighted by transmittance) produces a final 8-bit color value, the rendered pixel color, which is compared to the corresponding ground truth 8-bit pixel color to supervise NeRF training.

In contrast, path tracing assumes radiance values are expressed in linear color space.  
To relate NeRF and surface rendering in a physically meaningful way, they should ideally operate in a standard color space.  
To reconcile this difference, we train an HDR variant of NeRF, supervised with 32-bit HDR images directly rather than the standard 8-bit NeRF.  The resulting HDR NeRF produces a 3-channel radiance in 32-bit \textit{linear color space} at each sampled point.  For details regarding HDR data acquisition, preprocessing, and HDR NeRF implementation, see ~Appendix~\ref{app:hdr_details}.  As our focus here is to articulate the advantage of HDR NeRF in the context of the overall system, we forego a more general discussion of training NeRF in HDR and refer interested readers to \cite{mildenhall2022rawnerf} for a deep dive.  The HDR NeRF rendering equation can thus be written as:
$L_i(p, \omega) = \int_{0}^{t_f} T(t)\sigma(p_t) L_i(p_t, \omega) dt$
% \begin{align}
% \label{eq:nerf_rendering_hdr}
%     L_i(p, \omega) &= \int_{0}^{t_f} T(t)\sigma(p_t) L_i(p_t, \omega) dt
% \end{align}
where the transmittance term $T(t)$ remains unchanged.

With this simple adjustment, the NeRF equation can now directly relate to the surface rendering equation.  
If capturing HDR training data is impractical, one can still use standard (LDR) NeRF. 
Since the NeRF volume acts as the only light source, and the total energy is dissipative during the light transport, $L_i(p, \omega)$ will never exceed the NeRF volume's maximal radiance.  $\psi(\cdot)$ would then degenerate to an identity mapping such that $L(p, \omega) = r(p, \omega)$.  In other words, in many practical cases, reasonable visual results could still be obtained if standard NeRF is used with our system, despite the resulting inaccuracy in the light transport simulation.

\mypara{Estimating light sources with differentiable surface rendering.}
\agao{
Sampling light sources for computing the shadow pass requires an approximate representation of the light sources (emitters) of the scene.  Note that NeRF's volume rendering formulation bakes appearance into each point in the volume rather than simulating physically based light transport.  
To recover an explicit representation of light sources, we first reconstruct the scene's geometry as a neural SDF using \emph{MonoSDF}~\cite{Yu2022MonoSDF}, from which we extract an explicit mesh.  Then, we employ a differentiable path tracer, \emph{Mitsuba3}~\cite{Mitsuba3, Jakob2022DrJit}, to estimate a UV Emission Texture for the mesh.  We follow the general approach of \cite{nimierdavid2021material}, though we customize the optimization procedure since our goal is to estimate only the light sources, as opposed to a full BRDF estimation (more details about the optimization can be found in ~Appendix~\ref{app:mitsuba_details}).
}

\agao{
Once the light source estimation has converged, we prune faces whose emission falls below a threshold from the explicit mesh, which is necessary for efficiency, as most faces in the explicit mesh do not emit light.  The hybrid renderer then consumes the pruned mesh.
}
\begin{figure}
\centering
\begin{tabular}{@{}c@{\hspace{1mm}}c@{}}
    \includegraphics[width=0.5\linewidth]{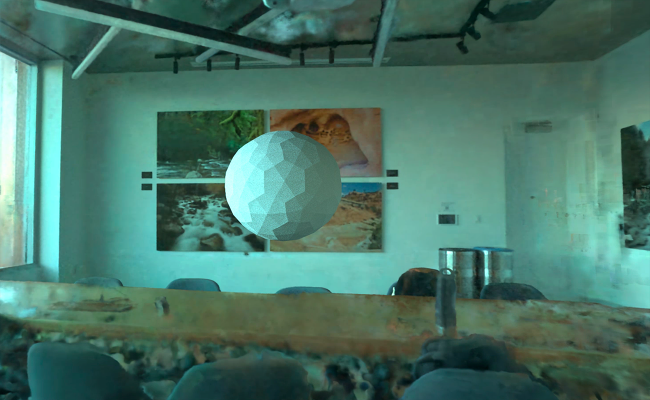} &
    \includegraphics[width=0.5\linewidth]{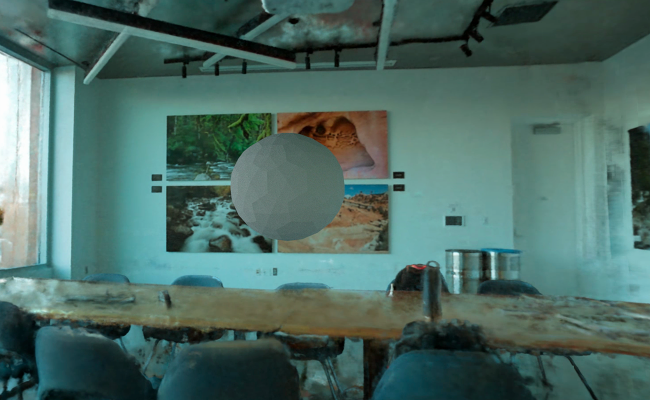} \\
    \small (a) HDR & \small (c) LDR  \\
    \includegraphics[width=0.5\linewidth]{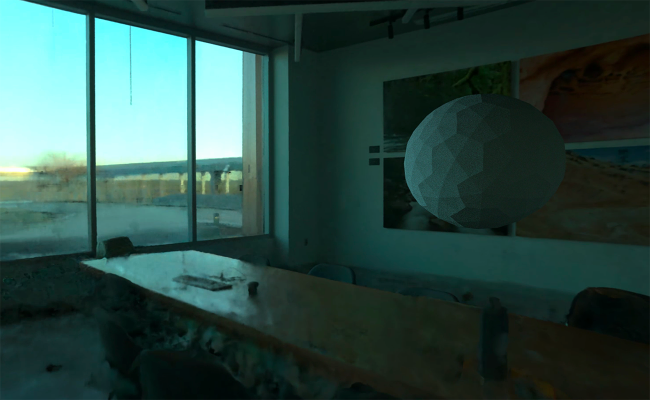} &
    \includegraphics[width=0.5\linewidth]{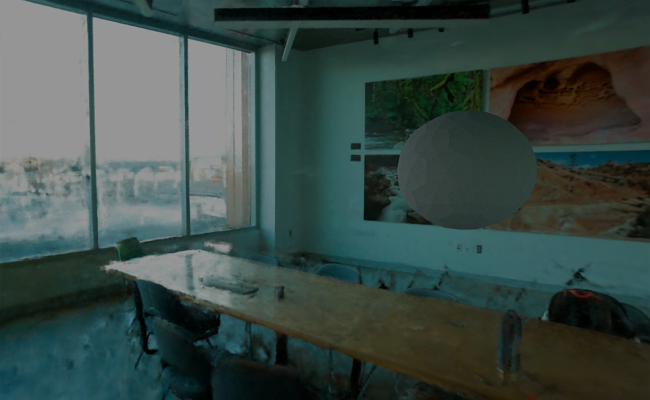} \\
    \small (b) HDR [-3.5 EV] & \small (d) LDR [-1.5 EV] \\
\end{tabular}
% \vspace{-1em}
\caption{\textbf{HDR Volumetric Radiance Map.} The diffuse sphere is rendered using our hybrid algorithm.  Images rendered using the NeRF trained in 32-bit HDR (a, b) achieve a higher level of lighting realism than those rendered with 8-bit LDR (c, d).}
\label{fig:comp_hdr}
% \vspace{-1em}
\end{figure}

\mypara{Shadow rays.}
\yiling{
% Accurate relighting and shadow effects on NeRF usually require extra information (e.g. varied lighting conditions~\cite{**}) and intrinsic decomposition to get the surface, light source, and material information, which is an ambiguous and time-consuming process.
We query additional rays during ray-marching to cast shadows on NeRF. For each sampled point $p_t$ in NeRF, we shoot a secondary ray from $p_t$ to the light source (see the following subsection for details on estimating lighting sources). 
If an inserted mesh blocks this ray, then this $p_t$ has a shadow mask $m(p_t)=1-r_{src}$, where $r_{src}$ is the intensity of the light source. Non-blocked pixels have $m_{shadow}=1$. The contribution of this point in Eq.~\ref{eq:nerf_rendering} is then $T(t)\sigma(p_t) m(p_t) r(p_t, \omega)$. Fig.~\ref{fig:shadow_ablation} shows that an inserted metal basket casts shadows on the NeRF desk.
}

\mypara{Hybrid Rendering Algorithm}
Based on the Light Transport Equations mentioned above, we note that both surface and NeRF rendering integrate the throughput $T(p)$ and radiance $r(p, \omega)=L_i(p,\omega)$. The differences are: (1) Surface rendering updates those values on discrete boundaries while NeRF accumulates them in the continuous space; (2) $T(p)$ and $r(p, \omega)$ are governed by the BSDF parameters in surface rendering, while by neural fields in NeRF. Therefore, we can alternate between the surface and NeRF rendering rules as they travel in space.
Algorithm~\ref{alg:render} is a summary of the hybrid rendering of NeRF and surface representations:
\begin{enumerate}
\setlength{\itemsep}{1mm}
\setlength{\parskip}{0pt}
\item We use Monte Carlo path tracing to sample the ray-surface-intersections $p_0\rightarrow p_1 \rightarrow ... \rightarrow p_n$, where $p_0$ is the camera center, and $p_n$ is the termination of the path. At the beginning of the path, initialize accumulated throughput $T(p_0)=1$ and radiance $r(p_0, \omega_0)=(0,0,0)$. The termination conditions will be discussed in (4).

\item \yiling{If shadows are needed, we estimate light source geometry and intensity $r_{src}$ with differentiable surface rendering.}

\item  For each ray segment $p_j\rightarrow p_{j+1}$, we use the ray-marching algorithm to sample and integrate the NeRF medium. \yiling{For the sampled points $p_t$ on the ray, we shoot a ray from $p_t$ to the light source (if any). If meshes block this point, set its \textit{shadow mask} to be $m(p_t)=1-r_{src}$ (or simply a constant close to 0), otherwise $m(p_t)=1$.} Then the throughput and \textit{shadow masked} radiance between surface intersections $p_j$ and $p_{j+1}$ can be computed as,
\begin{equation}
\label{eq:ray_marching_T}
   T'(p_{j+1}) = T(p_{j}) \cdot \exp\Big(-\int_{p_t}\sigma(p_t)dt\Big) 
\end{equation}
\vspace{-0.5em}
\begin{equation}
\label{eq:ray_marching_r}
\begin{aligned}
   L(p_{j+1}, \omega_j) =\ & L(p_j, \omega_j)\ +\\
   & \int_{p_t} T(t)\sigma(p_t)m(p_t) r(p_t, \omega) dt 
\end{aligned}
\end{equation}
where $p_t\in (p_j,p_{j+1}]$ and\\
$$T(t)= T(p_j)\cdot\exp\Big(-\int_{0}^t\sigma(p_s)ds\Big)$$
is also accumulated from $p_j$.

\item At the end of a ray segment, we reach the interface $p_{j+1}$ where the surface-rendering procedures occur. The direction $\omega_{j+1}$ of the next ray is determined by sampling the BSDF, and the weighted illumination and emitted light at this point are added to the radiance: 
\begin{equation}
\label{eq:path_r}
    L(p_{j+1}, \omega_{j+1}) = L(p_{j+1}, \omega_j)+T(p_j)L_e(p_{j+1},\omega_j)
\end{equation}
The throughput weight is updated as:
\begin{equation}
\label{eq:path_t}
    T(p_{j+1}) = \frac{T'(p_{j+1})\cdot f_s(p_{j+1}, \omega_{j}, \omega_{j+1})\abs{\cos(\theta_{j+1})}}{P}
\end{equation}
where $P$ is the scattering probability density function.

\item In the end, the ray terminates at $(p_{e}, \omega_e)$ if (1) it runs out of the scene; (2) current throughput $T(p_{e})$ is lower than a threshold; or (3) it meets the bounce limit.

\item As the rendering procedure is carried out over a linear 32-bit color space after the path tracing terminates for a given pixel, we can apply a nonlinear tone-mapping function, which we denote as $\psi$, to map from linear radiance to final sRGB color $r(p_{e}, \omega_e)$ which is more suitable for displaying on a monitor:
\begin{equation}
r(p_{e}, \omega_e) = \psi (L(p_{e}, \omega_e))
\end{equation}
\end{enumerate}

\begin{algorithm}
\caption{Hybrid Rendering Pipeline}\label{alg:render}
\begin{algorithmic}
\Require Meshes and pretrained NeRF of the scene.
% \Ensure $y = x^n$
\State \yiling{Estimate light sources with differentiable surface rendering. }
\For{each pixel $(u,v)$ in parallel}
    % \For{Random samples $\{1,...,n_{sample}\}$}
        \State Initialize $p_0$, $\omega_0$ based on $(u,v)$ and camera center.
        \State Set throughput $T(p_0)=1$.
        \State Set radiance $L(p_0, \omega_0)=(0,0,0)$. 
        \For{$j \in \{1,...,n_{bounces}\}$}
            \State Cast ray to find next intersection $p_j$ (\textcolor{blue}{Eqn.~\ref{eq:cast}}).
            \State March along ray $p_{j-1}\rightarrow p_j$.
            \State \yiling{Cast shadow rays to light sources.}
            \State Integrate $T'(p_j)$ and $L(p_{j}, \omega_{j-1})$ (\textcolor{blue}{Eqn.~\ref{eq:ray_marching_T}, ~\ref{eq:ray_marching_r}}).
            \State Sample BSDF at $p_j$ to get next ray direction $\omega_{j}$
            \State Update $T(p_j)$ and $L(p_{j}, \omega_{j})$ (\textcolor{blue}{Eqn.~\ref{eq:path_t}, ~\ref{eq:path_r}}).
            \State \textbf{break} if termination conditions satisfied.
        \EndFor
    \State \textit{(Path tracing endpoint denoted as $(p_{e}, \omega_e)$)}
    % \State \yiling{Compute the shadow mask $m_{shadow}$ and multiply $L$ by it.}
    \State Apply tone-mapping function $r = \psi (L(p_{e}, \omega_e))$.
    % \EndFor
    % \State Normalize the accumulated radiance for each pixel.
\EndFor
\end{algorithmic}
\end{algorithm}

% \begin{figure}
% \centering
% \begin{tabular}{@{}c@{\hspace{1mm}}c@{\hspace{1mm}}c@{}}
%     \includegraphics[width=0.32\linewidth]{figure/garden_ours.jpg} &
%     \includegraphics[width=0.32\linewidth]{figure/garden_nv.png} &
%     \includegraphics[width=0.32\linewidth]{figure/garden_ns.png} \\
%     \includegraphics[width=0.32\linewidth]{figure/bicycle_ours.jpg} &
%     \includegraphics[width=0.32\linewidth]{figure/bicycle_nv.png} &
%     \includegraphics[width=0.32\linewidth]{figure/biycycle_ns.png}  \\
%     % \includegraphics[width=0.5\linewidth]{figure/garden_ns.png} &
%     % \includegraphics[width=0.5\linewidth]{figure/biycycle_ns.png}  \\ 
%     \small (a) Ours & \small (b) NVDiffrec & \small (c) Nerfstudio
% \end{tabular}
% \vspace{-1em}
% \caption{
% \textbf{Rendering comparison.} 
% We insert a reflective metal ball into the Garden and Bicycle~\cite{barron2022mipnerf360}. 
% For this virtual object insertion comparison, we use NVDiffrec to extract the foreground mesh and then add render the scene with the synthetic ball using ray tracing. 
% Mesh object reconstruction from images remains challenging. 
% The table surface in the first row is noisy, and NVDiffRec fails to reconstruct the thin structure like the bicycle's wheel and the bench.
% In contrast, we can see the inserted ball through the gaps in the bench. 
% }
% \label{fig:comp_render}
% \vspace{-1em}
% \end{figure}

\begin{figure*}
\centering
\begin{tabular}{@{}c@{\hspace{1mm}}c@{\hspace{1mm}}c@{\hspace{1mm}}c@{}}
    \includegraphics[width=0.24\linewidth]{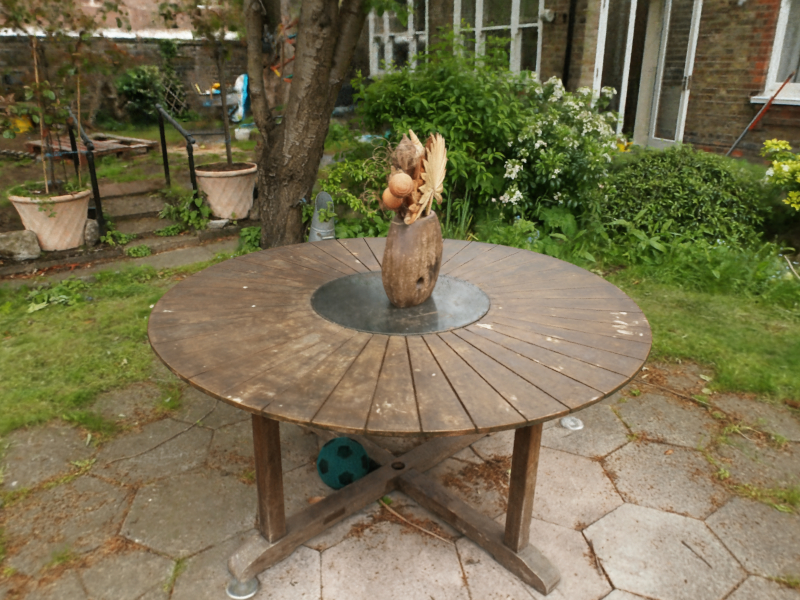} &
    \includegraphics[width=0.24\linewidth]{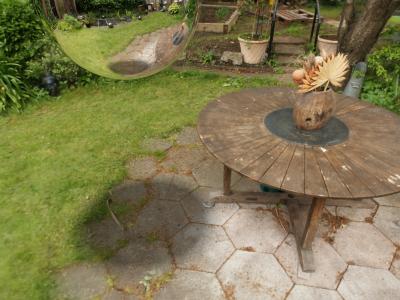} &
    \includegraphics[width=0.24\linewidth]{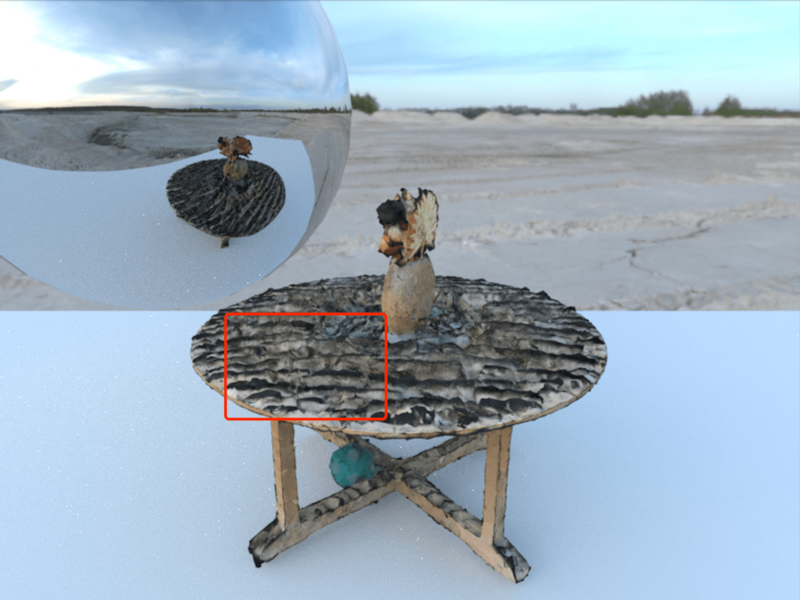} &
    \includegraphics[width=0.24\linewidth]{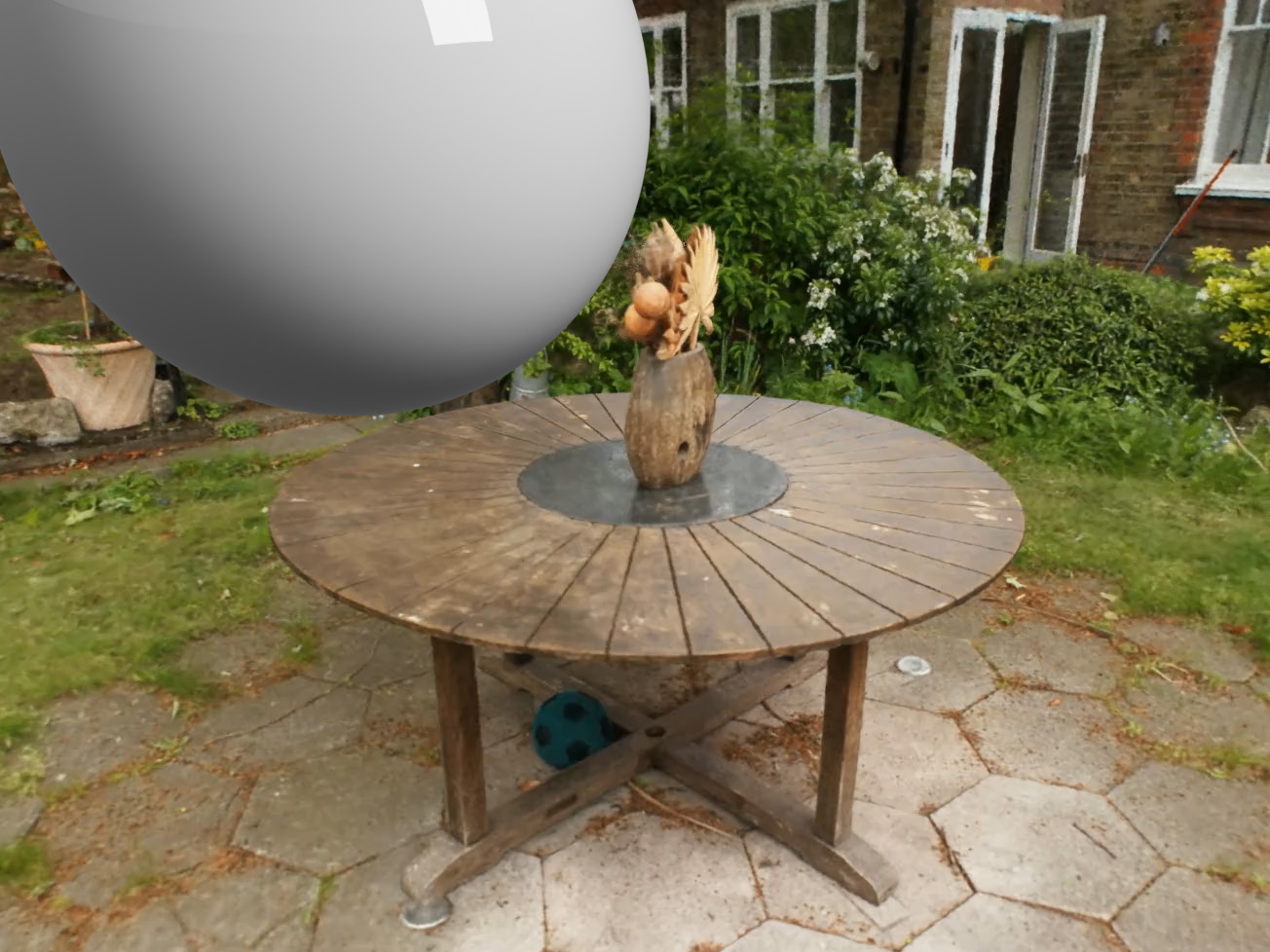} \\
    \includegraphics[width=0.24\linewidth]{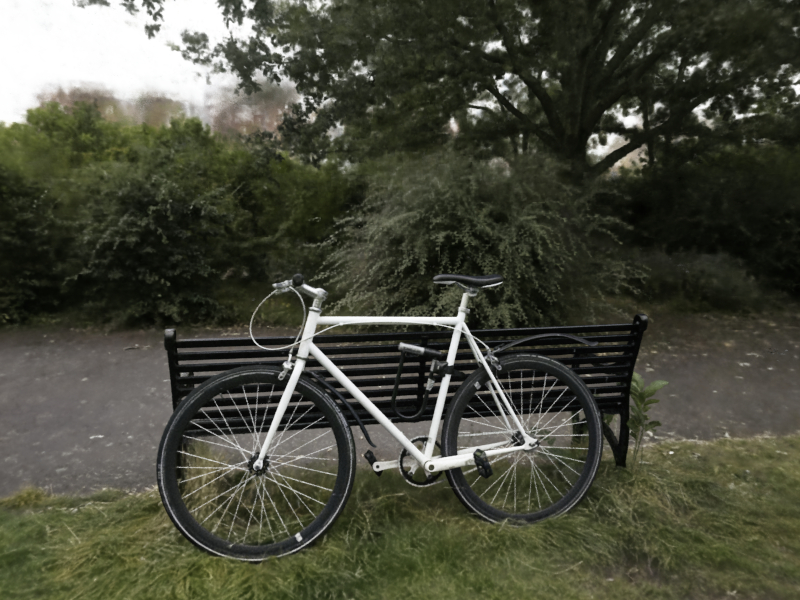}  &
    \includegraphics[width=0.24\linewidth]{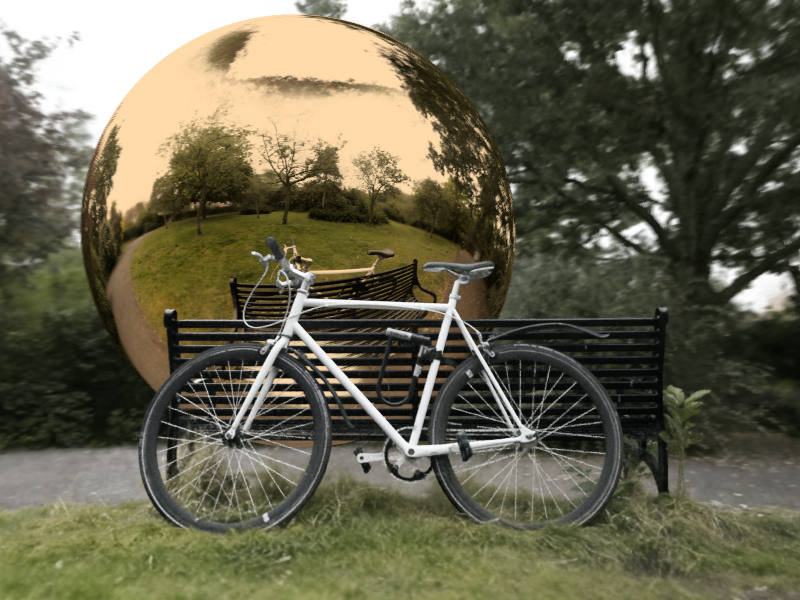} &
    \includegraphics[width=0.24\linewidth]{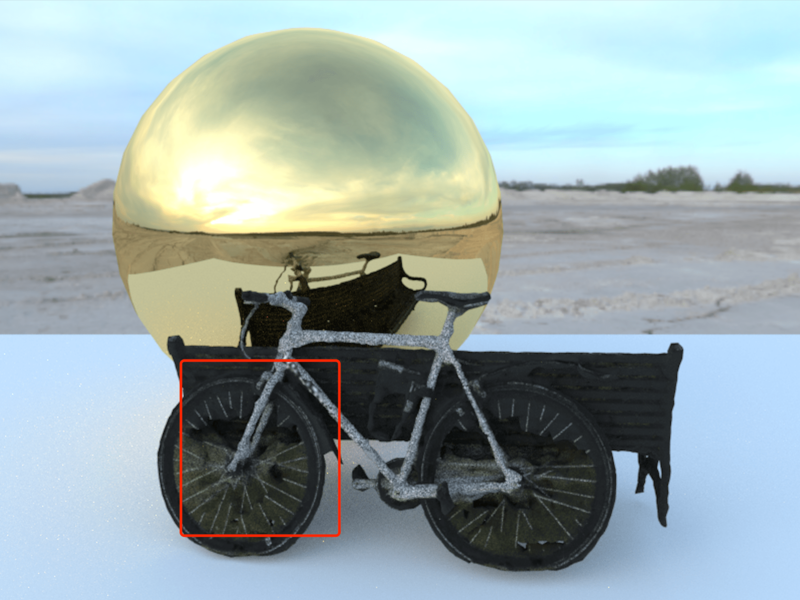} &
    \includegraphics[width=0.24\linewidth]{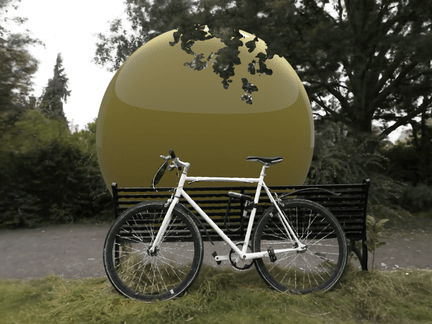}  \\
    \small (a) Original rendering & \small (b) Ours & \small (c) NVDiffrec & \small (d) Nerfstudio
\end{tabular}
% \vspace{-1em}
\caption{
\textbf{Rendering comparison for virtual object insertion.} 
We insert a reflective metal ball into the Garden and Bicycle scenes from the Mip-NeRF 360 dataset~\cite{barron2022mipnerf360}. (b) Our hybrid method produces results of superior visual quality with fewer artifacts. (c) We extract the foreground mesh using NVDiffRec, then insert the synthetic ball and render using ray tracing. Extracting an explicit mesh object results in noticeable artifacts such as the noisy table surface, and missing thin structures like the bicycle's wheel and the bench. (d) The 2D compositing workflow of Nerfstudio suffers from non-3D-aware occlusion masks and a limited ability to accommodate realistic interreflection.
}
\vspace{-1em}
\label{fig:comp_render}
\end{figure*}

\subsection{Simulation}
% \yiling{cite more simulation papers, add a high-level summary of the goals of this section}

We incorporate a dynamics simulator that supports rigid bodies, cloth, and deformable solids. Neural fields and meshes can be connected in the simulation pipeline by Signed Distance Fields (SDF) or reconstructed surface mesh. The SDF of the NeRF can be obtained in several ways during pre-processing. On the one hand, existing methods can directly learn the SDF, like NeuS~\cite{wang2021neus} and VolSDf~\cite{yariv2021volume}. On the other hand, it is common to set a threshold value for the density field and extract the surface mesh through Marching Cube~\cite{lorensen1987marching}. And the SDF can be converted from the mesh. The learned SDF has better quality but takes more time. We also implement an Instant-NGP version of NeuS (called NeuS-NGP) and accelerate the original code by more than 10 times. Users can make the trade-off depending on their needs.

% First, we assume the SDF of NeRF $s(\xx): \mathcal{R}^3\rightarrow\mathcal{R}$, $\xx \in \mathcal{R}^3$ is fixed (does not vary as a function of time) since we want to use NeRF to model a presumably static background. Meshes represent the other simulated objects with vertices $\{\vv_i\in\mathcal{R}^{n_i\times 3}\}_{i=1,2,..,k}$, where there are k meshes, and $n_i$ is the number of vertices of the i-th mesh.

We employ extended position-based dynamics (XPBD)~\cite{macklin2016xpbd,muller2020detailed} to simulate the objects during runtime. We choose this dynamics model because it is fast and can support various physical properties. 
Collision detection is performed by querying the SDF of all vertices. All of these queries can be computed efficiently in parallel on a GPU.
Given a detected collision from the SDF, we can also get the penetration depth and normal, which can be used to compute the contact forces.

\yiling{In some scenarios, NeRF can represent movable objects (e.g. a scene can be a composition of several NeRF objects~\cite{tang2022compressible}) instead of a static background. We can get the homogenous transformation $\tt\in\mathcal{R}^{4\times 4}$ of the NeRF from simulation in each time step, which is used to inverse transform the homogenous coordinates $\tt^{-1}\cdot \pp$~\cite{pumarola2020dnerf} when querying the color/density and sampling rays in the Instant-NGP hash grid. In Fig.~\ref{fig:game} (b), we control a ball to interact with a NeRF chair~\cite{mildenhall2020nerf} in real-time. The supplementary video further shows how the collision effect changes when the ball and chair have different relative mass and velocities.}
\subsection{Implementation Details}
% Our implementation is a proof-of-concept demonstration of the proposed hybrid rendering and simulation algorithm. 
% It aims to be interactive, user-friendly, and extendable, but its features are not as complete as classical rendering pipelines, as seen in Unity~\cite{haas2014history} or Blender~\cite{blender}. 
The entire pipeline employs CUDA backends for computation and Python interfaces for interaction. For rendering, NeRF is trained with the default configuration using Instant-NGP~\cite{mueller2022instant}. We also implement an instant-NGP~\cite{mueller2022instant} version of NeuS~\cite{wang2021neus} for efficient learning of the implicit SDF geometry. The path tracing algorithm is implemented using CUDA, embedded in Instant-NGP’s ray-marching procedures. We incorporate refractive, reflective, and Lambertian BSDF models. Physics simulation utilizes Warp~\cite{warp2022}, which just-in-time compiles Python code into CUDA kernels. The connection between rendering and simulation is facilitated by a Python interface using pybind11~\cite{pybind11}. Scene parameters can be easily created or modified through config files or Python APIs.

Our method can achieve a runtime of 1 to 40 frames per second, contingent upon the resolution, scene complexity, and dynamics. Figure~\ref{fig:game} demonstrates a real-time game on a laptop that has been developed within our pipeline. The code for rendering, simulation, and fast SDF learning will be released as open-source software.

\section{Experiments}
% \begin{figure*}
% \centering
% \begin{tabular}{@{}c@{\hspace{1mm}}c@{\hspace{1mm}}c@{\hspace{1mm}}c@{}}
%     \includegraphics[width=0.24\linewidth]{figure/ori_garden.jpg} &
%     \includegraphics[width=0.24\linewidth]{figure/garden_ours_shadow.jpg} &
%     \includegraphics[width=0.24\linewidth]{figure/garden_nv.png} &
%     \includegraphics[width=0.24\linewidth]{figure/garden_ns.png} \\
%     % \includegraphics[width=0.24\linewidth]{figure/ori_bicycle.jpg}  &
%     % \includegraphics[width=0.24\linewidth]{figure/bicycle_ours.jpg} &
%     % \includegraphics[width=0.24\linewidth]{figure/bicycle_nv.png} &
%     % \includegraphics[width=0.24\linewidth]{figure/biycycle_ns.png}  \\
%     % \includegraphics[width=0.5\linewidth]{figure/garden_ns.png} &
%     % \includegraphics[width=0.5\linewidth]{figure/biycycle_ns.png}  \\ 
%     \small (a) Original rendering & \small (b) Ours & \small (c) NVDiffrec & \small (d) Nerfstudio
% \end{tabular}
% \vspace{-1em}
% \caption{
% \textbf{Rendering comparison.} 
% We insert a reflective metal ball into the Garden and Bicycle~\cite{barron2022mipnerf360}. 
% For this virtual object insertion comparison, we use NVDiffrec to extract the foreground mesh and then add render the scene with the synthetic ball using ray tracing. 
% Mesh object reconstruction from images remains challenging. 
% The table surface in the first row is noisy, and NVDiffRec fails to reconstruct the thin structure like the bicycle's wheel and the bench.
% In contrast, we can see the inserted ball through the gaps in the bench. 
% }
% \vspace{-1em}
% \label{fig:comp_render}
% \end{figure*}

\subsection{Comparisons}
\mypara{Rendering Comparisons.}
In this section, we compare with other surface-based modeling and rendering methods in the virtual object insertion task. Given a set of images, the traditional graphics pipeline would first reconstruct the surface.%, set up the lighting and cameras, insert the virtual object, and then perform surface-based rendering like path tracing. 
However, the 3D reconstruction step will usually introduce tremendous noise and errors. Our technique can directly render the virtual object in the photorealistic 3D scene without meshing the entire scene. 

NVdiffrec~\cite{munkberg2021nvdiffrec} and IRON~\cite{zhang2022iron} are state-of-the-art textured mesh reconstruction methods. They combine neural fields and differentiable rendering methods to estimate the geometry and appearance of the objects from images. However, neither of them works on the full image because the topology of the background is too complex to optimize (e.g. the vegetation). We further provide the per-frame foreground mask to the comparison methods, and NVdiffRec can reconstruct foreground models. 
\begin{figure}
\centering
\begin{tabular}{@{}c@{\hspace{1mm}}c@{}}
%     \includegraphics[width=0.5\linewidth]{example-image} &
%     \includegraphics[width=0.5\linewidth]{example-image}  \\
%   \small (a) Rigid Body   & \small (b) Multibody    \\ 
    \includegraphics[width=0.5\linewidth]{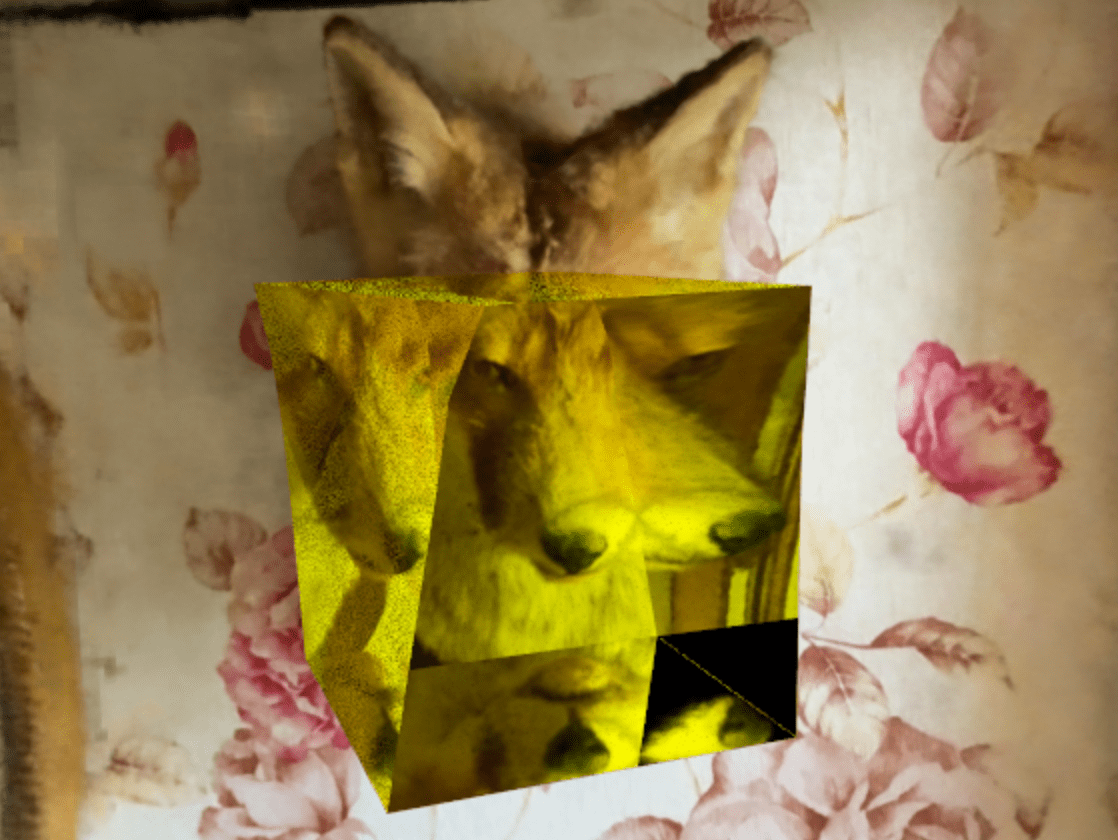} &
    \includegraphics[width=0.5\linewidth]{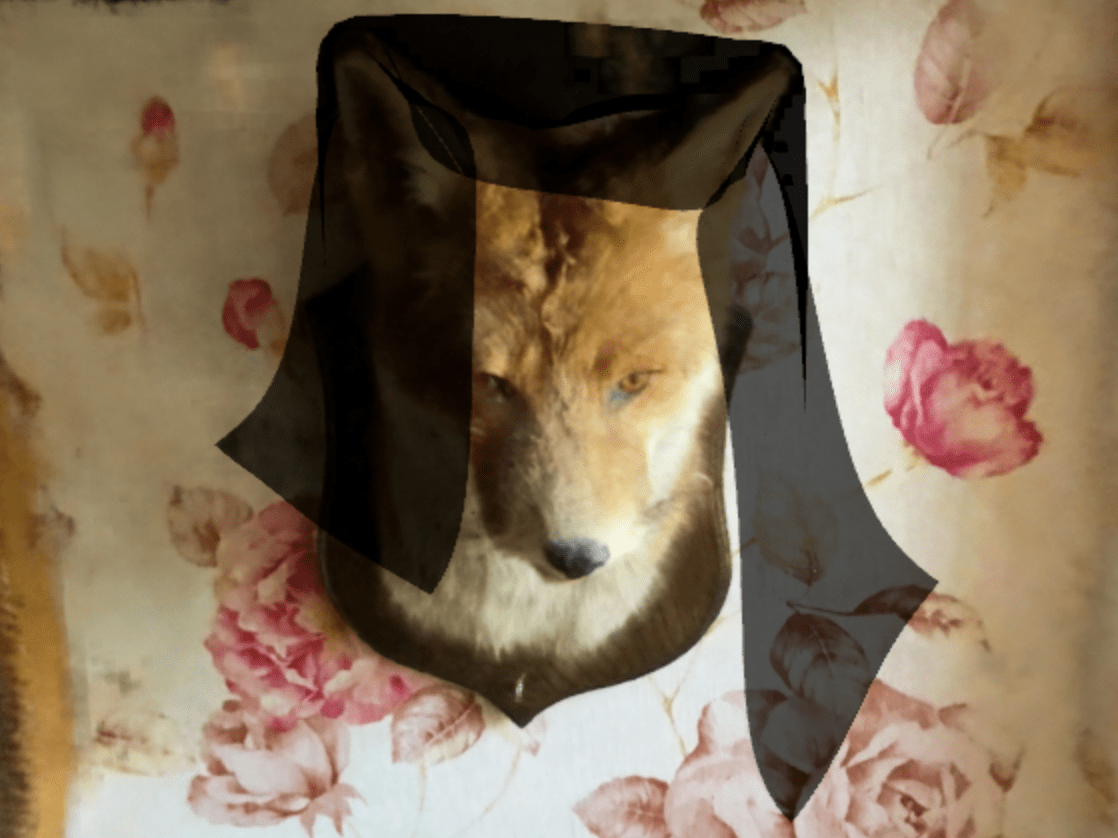}  \\
  \small (a) Soft body   & \small (b) Cloth   \\
\end{tabular}
% \vspace{-1em}
\caption{\textbf{Qualitative results for simulation.} In addition to rigid body simulation (Fig.~\ref{fig:game}), our method can also simulate soft bodies. (a) is a twisting Neo-Hookean FEM mesh. (b) is a thin shell cloth covering the fox. Please see the supplementary video for more simulation results.  }
\label{fig:qual_soft}
% \vspace*{-1em}
\end{figure}

In Figure~\ref{fig:comp_render} (c), we import the extracted mesh into Blender with a metallic ball and environment map, then render the scene with Cycles, a physically-based render. Both (b) ours and (c) NVDiffRec can model the reflection on the ball, but our object has better surface quality.
% In the first row, our table is more smooth and more realistic. In the second row, the mesh in (b) fails to model the structure of the bench and spokes, while our method can faithfully visualize the structure of the bench and wheels.

We also compare to Nerfstudio~\cite{nerfstudio}, a software library that allows users to train their own NeRF models. The user can export a camera path moving through the trained NeRF scene and a coarsely reconstructed mesh and import these into Blender~\cite{blender} using the Nerfstudio Blender plugin.  The user may then create their 3D content directly in Blender, using the imported camera path and mesh as a reference, and then render their 3D content, which is then composited in 2D over the rendered NeRF trajectory.  %In Figure~\ref{fig:comp_render} (d), we show that while reflections of the environment may be imitated by employing a reasonable HDRI environment map in Blender, this approach cannot produce high-quality reflections of the true scene.  
However, while the coarse mesh may be used to generate a visibility mask to produce occlusion, this approach is limited in that the coarsely reconstructed mesh is likely not accurate enough to provide clean-looking occlusion results, more complex occlusion situations such as occluding objects at multiple layers of depth will be challenging to generate individual visibility masks, and this requires manual effort to composite the results.
\yiling{
Appendix~\ref{app:details_comparison} includes some more quantitative comparisons with those methods. Moreover, LumaAI recently released a closed-source UE plugin, and we run a comparison against it.
}

% In summary, with our hybrid rendering technique, users bypass the 3D surface reconstruction step and are able to directly render their 3D mesh with the real-world background. 

% \cite{munkberg2021nvdiffrec}
% \cite{zhang2022iron}

% \begin{figure}
% \centering
% \begin{tabular}{@{}c@{\hspace{1mm}}c@{\hspace{1mm}}c@{}}
%     \includegraphics[width=0.32\linewidth]{example-image} &
%     \includegraphics[width=0.32\linewidth]{example-image} &
%     \includegraphics[width=0.32\linewidth]{example-image} \\
%   \small (a) Gripper   & \small (b) Octopus    & \small (c) Fish   
% \end{tabular}
% \vspace{-0.5em}
% \caption{}
% \vspace{-1.5em}
% \label{fig:motion}
% \end{figure}

\mypara{Rendering with HDR NeRF.}
We provide a qualitative comparison between rendering using the standard (LDR) NeRF and its HDR counterpart, shown in Figure~\ref{fig:comp_hdr}.  Notice that in the images rendered with the HDR model, the lighting cast from the environment onto the mesh appears much more faithful to the scene's true intensity (and therefore directionality).  This is also not surprising, considering the longstanding use of High Dynamic Range Image-Based Lighting in the traditional graphics pipeline \cite{debevec_hdr_ibl}.  HDR images are often created by recovering the unknown nonlinear tone-mapping function from a series of bracketed-exposure LDR images with known exposure duration, then using the inverse function to map the images back to linear color space, and merging them into a single 32-bit result, as introduced by Debevec and Malik \cite{debevec1997}.  As a result, HDR images are much better suited to capture a scene's full range of absolute and relative radiance values and avoid highly lossy clipping, which can be especially problematic in very bright parts of an image.  HDR NeRF may therefore be interpreted as a volumetric HDR lighting map.

Importantly, our hybrid rendering algorithm allows us to utilize such an HDR volumetric radiance map fully.  As an HDR radiance map is particularly useful for representing the indirect bounce lighting of the scene, future work on extending such learned lighting models to direct lighting (e.g. directional or point sources that cast hard shadows) would be a promising direction.  Our hybrid rendering algorithm is crucial to enable such an investigation.

\mypara{Simulation Comparisons.}
Besides static scenes, our pipeline can also simulate dynamic scenes with SDF-based contact handling. Appendix~\ref{app:density} shows that the SDF-based representation has better collision handling than vanilla NeRF density fields. We also compare with NeRF-based simulation~\cite{Cleac2022Differentiable,qiao2022neuphysics} in Appendix~\ref{app:comp_nerf_sim} and ours achieves better performance.

\begin{figure}
\centering
\begin{tabular}{@{}c@{\hspace{1mm}}c@{}}
    \includegraphics[width=0.5\linewidth]{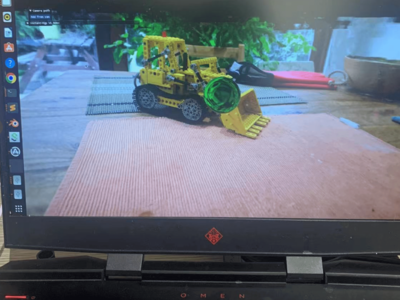} &
    \includegraphics[width=0.5\linewidth]{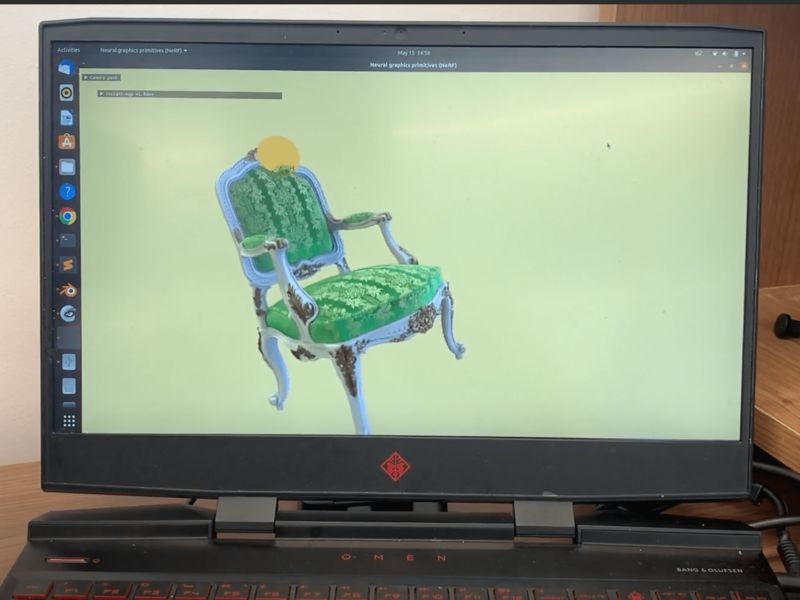}  \\
    \small (a) Static NeRF background & \small (b) NeRF (chair) as dynamic object 
\end{tabular}
% \vspace{-1em}
\caption{\textbf{Real-time photorealistic gaming on a laptop.} Our rendering and simulation engine can be interactive 
and run in real-time on a laptop with an NVIDIA GeForce RTX 2070 Max-Q GPU. 
In this example, a user can control the motion of the synthetic glass ball and interact with the background (collision and light effect). 
} 
\label{fig:game}
% \vspace{-1em}
\end{figure}

% We compare the simulation quality with different collision geometry. In Figure~\ref{fig:comp_simulation}, we drop four balls above a messy counter top~\cite{barron2022mipnerf360}. In this simulation scene, (a) models the collision geometry using learned SDF by our NeuS-NGP and (b) directly uses Marching-Cube mesh from instant-NGP density fields (with resolution $256\times256\times256$). 
% Two of the balls in (b) `sink' into the countertop, which is unrealistic.  
% In contrast, our SDF-based simulation in (a) can correctly simulate the interaction between the balls and the background objects. More dynamics results can be found in our supplementary video. 

\subsection{Performance}
In this section, we will show how our method can be used in photorealistic real-time gaming and physically-based simulation. A detailed profiling of our rendering and simulation modules can also be found in Appendix~\ref{app:runtime}.

\mypara{Photorealistic Real-time Gaming.}
In addition to photorealistic rendering, our pipeline is fast, aiming to serve as a real-time neural-fields game engine. As shown in Figure~\ref{fig:game}, we have implemented an interactive game where players can control the ball's motion using a keyboard and adjust camera viewing angles using a mouse. The background scene, excluding the green ball, is modeled by NeRF. The ball can have contact with the table and bulldozer. Players can also observe the bending of rays as they pass through the refractive glass ball. Supplementary material includes a recording of the real-time game. The game runs on a laptop with an NVIDIA GeForce RTX 2070 Max-Q GPU. Through our pipeline, game developers can seamlessly integrate animatable objects with photorealistic NeRFs.

\mypara{Qualitative Results.}
Our method can simulate different types of dynamics with the support of Warp~\cite{warp2022}. 
% In addition to rigid body dynamics as shown in Figure~\ref{fig:comp_simulation}, \ref{fig:game}, our method can also support soft solid and thin shell cloth. 
In Figure~\ref{fig:qual_soft}, the background fox is modeled as NeRF. (a) is a cube modeled by Neo-Hookean FEM mesh, where we can also see the changing light effect as it is twisted. (b) is a piece of cloth falling down to the fox. Our pipeline can handle the collision and light effects of such thin shells.

% \subsubsection{Texture materials}

% \subsubsection{Simulation dynamics}

\subsection{Applications}
Our methods could be applied to many situations, where NeRFs can improve the realism of synthetic scenes.

\begin{figure}
\centering
\begin{tabular}{@{}c@{\hspace{1mm}}c@{}}
    \includegraphics[width=0.5\linewidth]{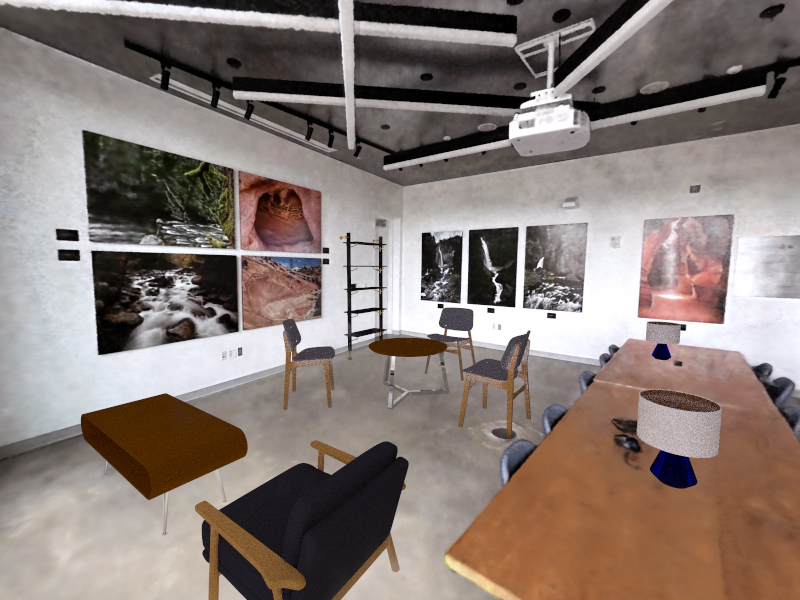}  &
    \includegraphics[width=0.5\linewidth]{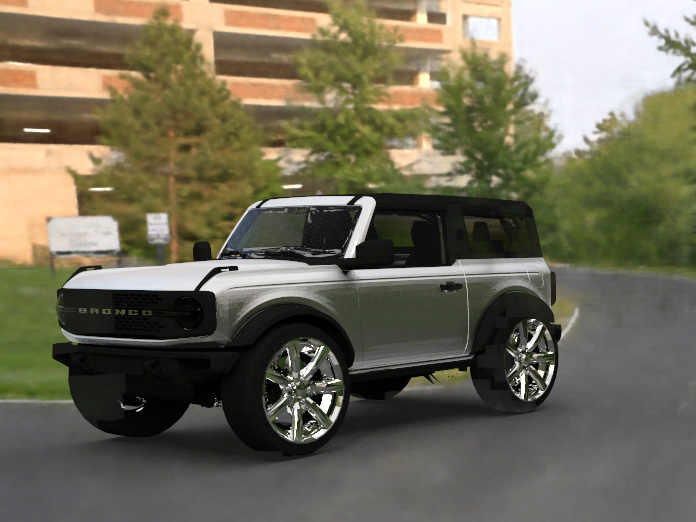} \\
  \small (a) Room layout design   & \small (b) Driving simulator  \\  \\
  % \vspace{0.5em}
    \includegraphics[width=0.5\linewidth]{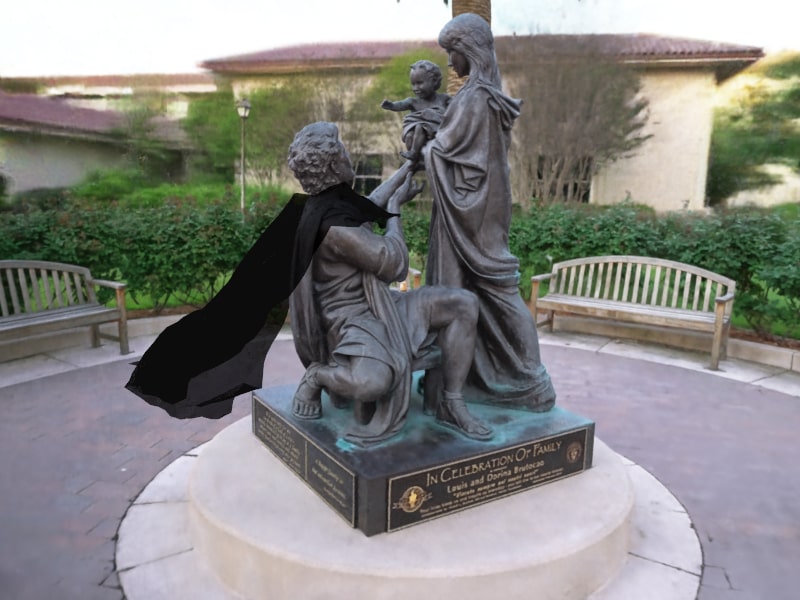} &
    \includegraphics[width=0.5\linewidth]{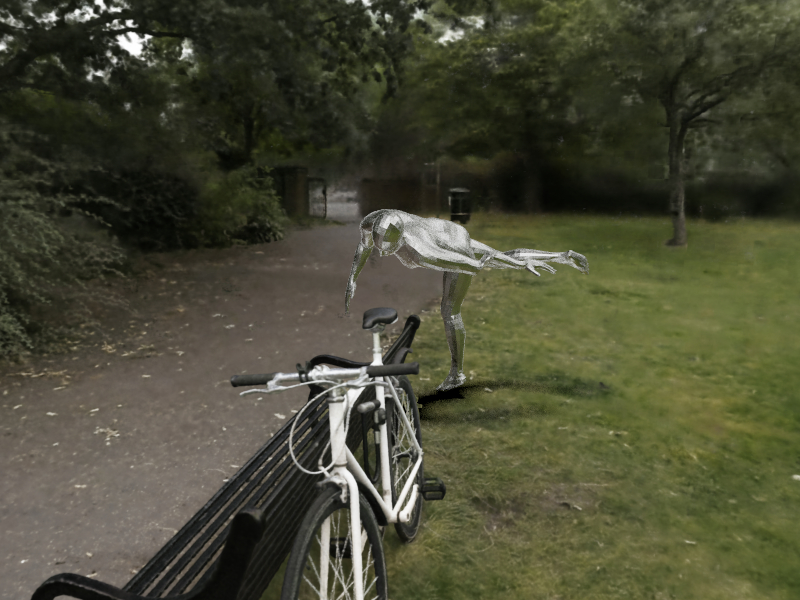}  \\
  \small (c) Virtual try-on   & \small (d) Digital human   \\
\end{tabular}
% \vspace{-0.75em}
\caption{\textbf{Application of mixing NeRF with meshes.} Our method can be used for realistic driving simulation, room layout design, virtual try-on, and digital humans. }
% \vspace{-1em}
\label{fig:app}
\end{figure}

\mypara{Driving simulation} is important for developing, training, and testing autonomous driving systems. With the large number of images captured around roads, people can train NeRF for street views~\cite{tancik2022block}. With our method, people can set up the driving simulation inside those Photorealistic NeRFs (see Figure~\ref{fig:app} (a)) and insert synthetic vehicles~\cite{car1,car2}. Such realistic virtual environments can help minimize the sim-to-real gap in self-driving cars. 

\mypara{Room layout design} can help users design their homes and purchase furniture. After taking pictures and building a NeRF model for their room, customers can shop
furniture~\cite{fur1,fur3} 
% \cite{fura,furb,furc} % I don't know, these three references lead to error
virtually and design the room layout as shown in Figure~\ref{fig:app} (b).

\mypara{Virtual try-on} using our methods can dynamically simulate the cloth on a human body captured by NeRF. In Figure~\ref{fig:app} (c), we place a cloak on the human body and simulate how it swings in the wind.

\mypara{Digital human} applications are one of the key interests in VR/AR and the metaverse. With our methods, users can easily collect and build their virtual world by NeRF and then render their human-body model in that scene. This could be useful for movie making, webcasting, virtual performance, cyber-tourism, etc. Figure~\ref{fig:app} (d) renders a futuristic `mercury man'~\cite{SMPL:2015} jumping in the park using NeRF. % Everyone can be a dancer in the stage made of NeRF!

\section{Conclusion}
% \agao{mention this Limitation? -- radiance of the environment affects rendering of the mesh, but the opposite isn't true -- radiance emitted from meshes does not affect the environment volumetric radiance}

In summary, motivated by integrating NeRF into the traditional graphics pipeline, our hybrid rendering method can render dynamically changing meshes in a photo-realistic NeRF environment, without costly surface reconstruction. We also equip the resulting renderer with a simulator, making it suitable as a real-time NeRF-based game engine.

% To integrate mesh-based and volume rendering, we start with their rendering equations. Although surface path tracing is integrated over discrete path space and NeRF is integrated over a continuous ray, the radiance contribution of each surface-ray interface and NeRF point is independent and additive, as long as there is a well-defined compatible weight. We then set rules for the weight and radiance update, which can then take into account both surface and neural fields when computing a Monte Carlo-sampled ray. Moreover, we design a neural-SDF-mesh simulator with Warp~\cite{warp2022}, which supports rigid bodies, soft bodies, and cloth dynamics. The entire pipeline can interact with users in real-time.

% While baking in the attenuation and view dependence makes the volume rendering more efficient, it also loses some of the raw information of the scene. In order to reconstruct the scene for relighting and editing, one must perform intrinsic decomposition on NeRF~\cite{srinivasan2021nerv}. Solving the inverse rendering problem is not within the scope of this work. Rather, we focus on providing a way to incorporate existing vanilla NeRF (e.g. a pre-trained Instant-NGP) into surface-based rendering and simulation. More complicated features like shadows and global illumination are left for future work.
There are some limitations in this work. (1) The currently implemented renderer cannot cast shadows and illumination on NeRF points. Decomposing NeRFs can make the relighting more realistic. (2) Our renderer offers basic, essential functions; support for environment maps, UV maps, and image textures for higher rendering quality can be a natural extension. (3) Additional interfaces can also enable users to take advantage of more mature infrastructures if integrated into more widely-used platforms, e.g. Blender, Unreal, etc. (4) We can further improve the runtime performance and integrate the pipeline with larger-scale NeRFs.

\mypara{Acknowledgements.} This research is supported in part by Dr. Barry Mersky and Capital One E-Nnovate Endowed Professorships, and ARL Cooperative Agreement W911NF2120076. Yi-Ling would also like to thank the support from Meta Fellowship and Dr. Chunsheng Hu's assistance in drawing the diagram.

% \clearpage
{\small
\bibliographystyle{ieee_fullname}
\bibliography{main}
}

%%%%%%%%%%%%%%%%%%%%%%%%%%%%%%%%%%%%%%%%%%%%%%%%%%%%%%%%%%%%%%%%%%%%%%%%%%%%%%%
%%%%%%%%%%%%%%%%%%%%%%%%%%%%%%%%%%%%%%%%%%%%%%%%%%%%%%%%%%%%%%%%%%%%%%%%%%%%%%%
% APPENDIX
%%%%%%%%%%%%%%%%%%%%%%%%%%%%%%%%%%%%%%%%%%%%%%%%%%%%%%%%%%%%%%%%%%%%%%%%%%%%%%%
%%%%%%%%%%%%%%%%%%%%%%%%%%%%%%%%%%%%%%%%%%%%%%%%%%%%%%%%%%%%%%%%%%%%%%%%%%%%%%%
\newpage
\appendix
\onecolumn
\section{Runtime and Memory}
\label{app:runtime}
To break down the time and memory cost of our method, Figure~\ref{fig:number} profiles the simulation and rendering modules as the number of synthetic objects increases. In the test scene, we throw $n\in\{1,10,100,1000,10000\}$ balls onto a Fox~\cite{mueller2022instant} represented by NeRF. The rendering resolution is $900\times 450$.

As seen in the figure, the simulation and rendering memory stay constant at a low level (0.5 GB and 1.0 GB, respectively) throughout the entire experiment. The run time for both modules scales practically linearly w.r.t. the number of objects.

% \begin{figure}
% \centering
% \vspace*{-1.5em}
% % \label{fig:resolution}
% \includegraphics[width=0.4\linewidth]{figure/plt_resolution.png}
% % \vspace{-5mm}
% % \caption{Scale the resolution. }
% % \vspace{5mm}
% \vspace*{-2em}
% \end{figure}
Moreover,
the simulation module is independent of the resolution, and the rendering time scales linearly w.r.t. the number of pixels (see left). This pipeline can run in real time (20 FPS) at $600\times 300$ resolution.

\section{Comparison with density fields}
\label{app:density}
We compare the simulation quality with different collision geometry. In Figure~\ref{fig:comp_simulation}, we drop four balls above a messy counter top~\cite{barron2022mipnerf360}. In this simulation scene, (a) models the collision geometry using learned SDF by our NeuS-NGP and (b) directly uses Marching-Cube mesh from instant-NGP density fields (with resolution $256\times256\times256$). 
Two of the balls in (b) `sink' into the countertop, which is unrealistic.  
In contrast, our SDF-based simulation in (a) can correctly simulate the interaction between the balls and the background objects. More dynamics results can be found in our supplementary video. 

\section{Comparison with other NeRF-based simulation.}
\label{app:comp_nerf_sim}
There are two recent works~\cite{Cleac2022Differentiable,qiao2022neuphysics} that can simulate in neural fields. \cite{Cleac2022Differentiable} establish a collision model based on the density field. In our simulation, we found that the density fields are usually noisy and inadequate to model a surface well for accurate contact processing. 
% Figure~\ref{fig:comp_simulation} (d) is a surface reconstructed from Instant-NGP density field, which might not be applicable for simulation. 
NeuPhysics\cite{qiao2022neuphysics} aims at differentiable simulation and rendering, and it extracts hexahedra mesh from learned SDF~\cite{wang2021neus} and simulates the mesh using \cite{du2021diffpd}. NeuPhysics can only simulate existing NeRF objects instead of synthetic objects.
Table~\ref{tab:speed_comparison} shows that our method runs at least an order of magnitude faster when simulating a bouncing ball.

\section{Light Source Estimation - Implementation Details}
\label{app:mitsuba_details}
\subsection{Geometry Reconstruction}
\agao{
Given our choice to represent light sources as area lights defined on an explicit mesh, we first need to reconstruct the geometry of the scene.  We choose to use MonoSDF \cite{Yu2022MonoSDF}, which leverages monocular geometric priors (depth and normal estimation) in the neural implicit surface reconstruction process.  Importantly, these priors especially help in reconstructing surfaces that have only a low degree of visual texture, which is very common for walls and flat tabletop surfaces.  For differentiable surface rendering, having reasonably accurate geometry is critical for achieving good convergence, as noisy geometry will tend to bias the result to bad local minima.  Empirically, MonoSDF achieves sufficiently accurate geometry for this purpose.
}
\agao{
Once we have an optimized signed distance field, we convert it to an explicit mesh using Marching Cubes, then UV unwrap the mesh to obtain a UV texture map, using SDFStudio \cite{Yu2022SDFStudio, nerfstudio}.  This step also results in a UV texture map whose values are determined by querying the underlying MonoSDF appearance model, which we use as an albedo UV texture map.  The UV-unwrapped mesh is the input to the differentiable surface render, described below.
}
\subsection{Differentiable Surface Rendering}
\agao{
We implement our differentiable surface rendering optimization using Mitsuba3 \cite{Mitsuba3}, which is built on Dr. Jit \cite{Jakob2022DrJit}, a just-in-time compiler that is specialized for rendering use cases.  Our inverse rendering procedure is generally based on \cite{nimierdavid2021material}, though our specific implementation details are as follows.  The UV-unwrapped scene mesh is assigned a Principled BSDF material \cite{Burley2012PhysicallyBasedSA}, as well as an emission UV texture map, which constitute the inputs to the differentiable renderer.  The emission UV texture map, $T_{emission}$, with dimensions $(h_{tx}, w_{tx}, c_{tx})$, is the variable that we are interested in optimizing.  It is initialized uniformly with near-zero values.  The other BSDF parameters remain fixed (albedo texture map, specular transmission map with default value 1.0, and roughness map with default value 0.5).  For rendering, we use a Path-Replay Backpropagation (PRB) integrator \cite{Vicini2021PathReplay}, and limit the light transport simulation to a maximum depth of 3 (2 bounces).  This stands in contrast to differentiable rasterization approaches to inverse rendering, which are limited to a maximum depth of 2 (1 bounce), and are therefore unable to account for indirect lighting in the optimization.  We also directly render emitters (as opposed to only emitting light onto surfaces without being directly visible).
}

\begin{figure}
\centering
\begin{tabular}{@{}c@{\hspace{2mm}}c@{\hspace{2mm}}c@{}}
    \includegraphics[width=0.3\linewidth]{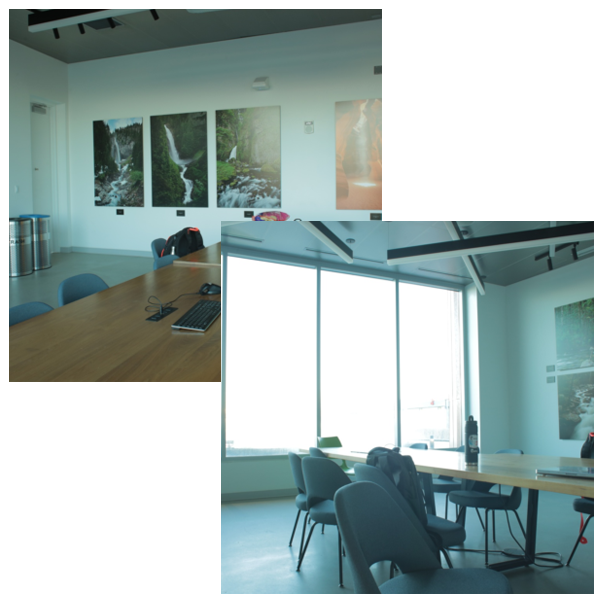} &
    \includegraphics[width=0.335\linewidth]{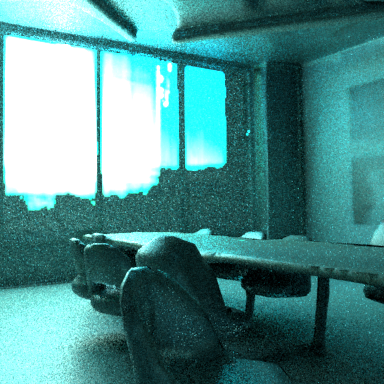} &
    \includegraphics[width=0.335\linewidth]{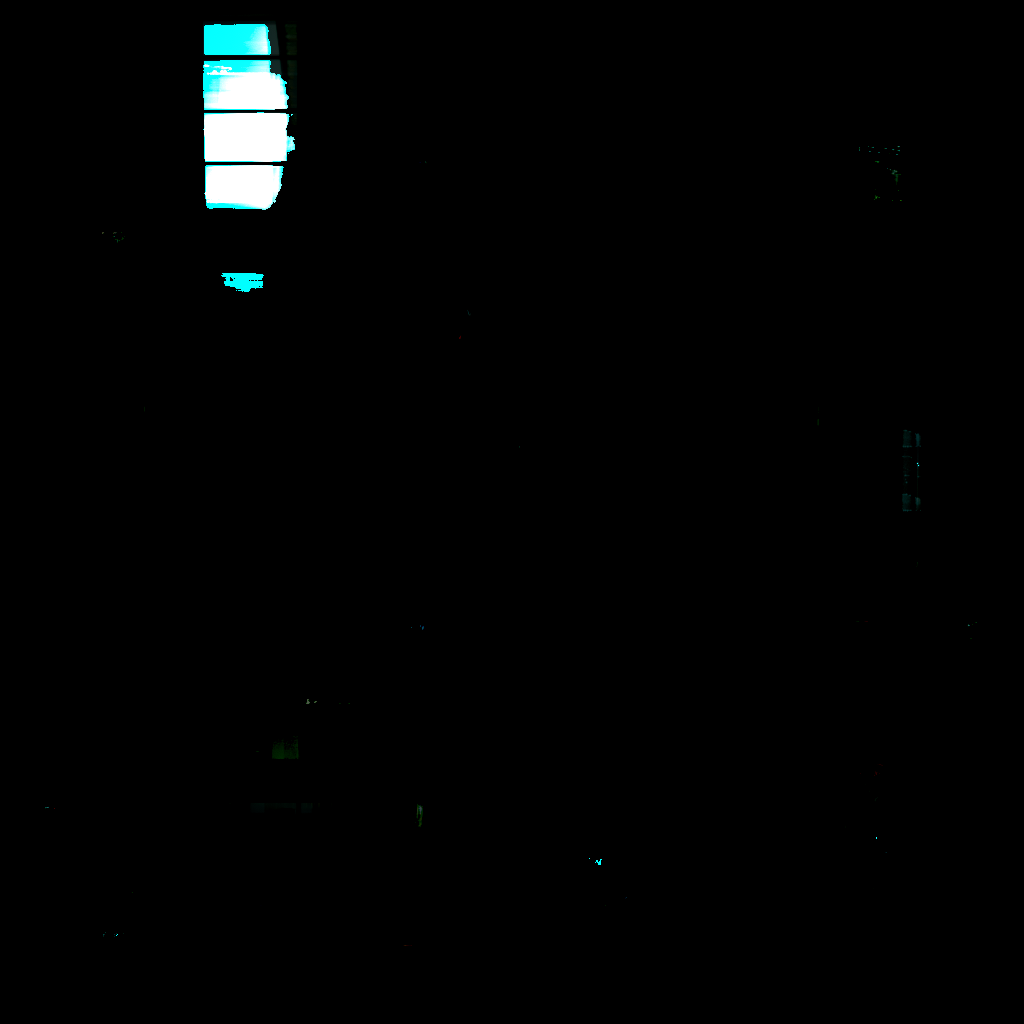} \\
    \small (a) Ground Truth Images & \small (b) Resulting image rendered with Mitsuba3 & \small (c) Optimized Emission UV Texture Map \\
\end{tabular}
\vspace{-2mm}
\caption{\textbf{Optimizing Emission UV Texture Map.} We optimize the (c) Emission UV Texture Map by minimizing a rendering loss objective between (a) ground truth images and (b) images rendered with Mitsuba3, a differentiable surface rendering engine.}
\label{fig:diffrender}
\end{figure}

\agao{
Given a set of ground-truth 32-bit HDR images $\{I_1, \cdots, I_n\}$ with dimensions $(h_{im}, w_{im}, c_{im})$ and known camera poses, our primary objective is to minimize the rendering loss (mean-squared error) between the ground truth images, and images rendered from identical camera poses.  Since the variables that we are optimizing are the emission UV texture map values, we also find that applying uniform L1 regularization to the optimization variable corresponds well to the inductive bias that the majority of the scene does not emit light, and therefore the emission UV texture map should be encouraged to be sparse.  For each optimization step $i \in [1, n]$ in a given optimization epoch, our objective is therefore to minimize the following, where $\alpha$ is a tunable hyperparameter: $$minimize\ \ \frac{1}{(h_{im} * w_{im} * c_{im})} \|I_i - \hat I_i\|^2 + \frac{\alpha}{(h_{tx} * w_{tx} * c_{tx})}|T_{emission}|$$
}
\agao{
In addition to the rendering loss and L1 regularization in texture space, we also find that periodically clipping low values and boosting large values in the emission texture space encourages convergence toward correct results.  The motivation is that periodically during the optimization procedure (according to a tunable cadence, where we assume a default period of every 2 epochs), emission values below a given brightness threshold (also a tunable hyperparameter, with a default value of 0.2), are clipped to 0.  By assuming that very dim values are not light sources, we avoid local minima.  At the same time, values that are above this threshold are likely to be light sources, and therefore, we boost these values, in order to accelerate the optimization, since extreme emission values in the 32-bit linear color space may be arbitrarily larger than 1.  This part of the optimization should be self-correcting: values that are boosted when they shouldn't be will then be brought down in order to satisfy the rendering loss objective.  See Figure~\ref{fig:diffrender} for an example of the converged result of optimizing the emission texture map.
}
\agao{
Finally, once the optimization has converged, we post process the mesh by pruning off any triangular faces whose vertices all fall below a threshold, by looking up the brightness of each vertex in the emission UV texture map, according to that vertex's associated UV coordinates.  This results in a greatly reduced mesh that only contains faces that should emit light onto the rest of the scene, and can therefore be used as an area light for computing the shadow pass of our Hybrid Renderer.
}

\section{HDR Rendering Experiment Details}
\label{app:hdr_details}
\subsection{Data acquisition}
To train the HDR NeRF, we first construct a set of 32-bit HDR images of a scene, where each image is analagous to a single LDR image in the baseline NeRF formulation.  To create a single one of these HDR images, we shoot a bracketed series of LDR exposures that consists of several images (7 in our experiments) captured from a fixed camera pose.  We use a single Canon 5D MKIII with a 28mm lens, and set the camera to record images in 8-bit JPEG format.  Specifically, the camera is mounted to a tripod to minimize any minute difference in camera pose between each of the exposures in the bracket, and to minimize any optical differences between the images (e.g. depth of field, bloom, etc.) apart from raw exposure value, we only adjust the exposure duration (i.e. shutter speed) between images, while the aperture remains fixed.  The \textit{relative} difference in exposure between any consecutive images in the bracket is kept constant, though the magnitude of this difference depends on the degrees of dynamic range of the scene being captured; ideally, within the range of exposures from darkest to brightest, every pixel of the image should be ``well-exposed" in at least one of the images.  We generally captured 7 images in the bracketed exposure, with 1-2 stops difference between each image.  

\subsection{Data preprocessing}
\begin{figure}
\centering
    \includegraphics[width=0.4\linewidth]{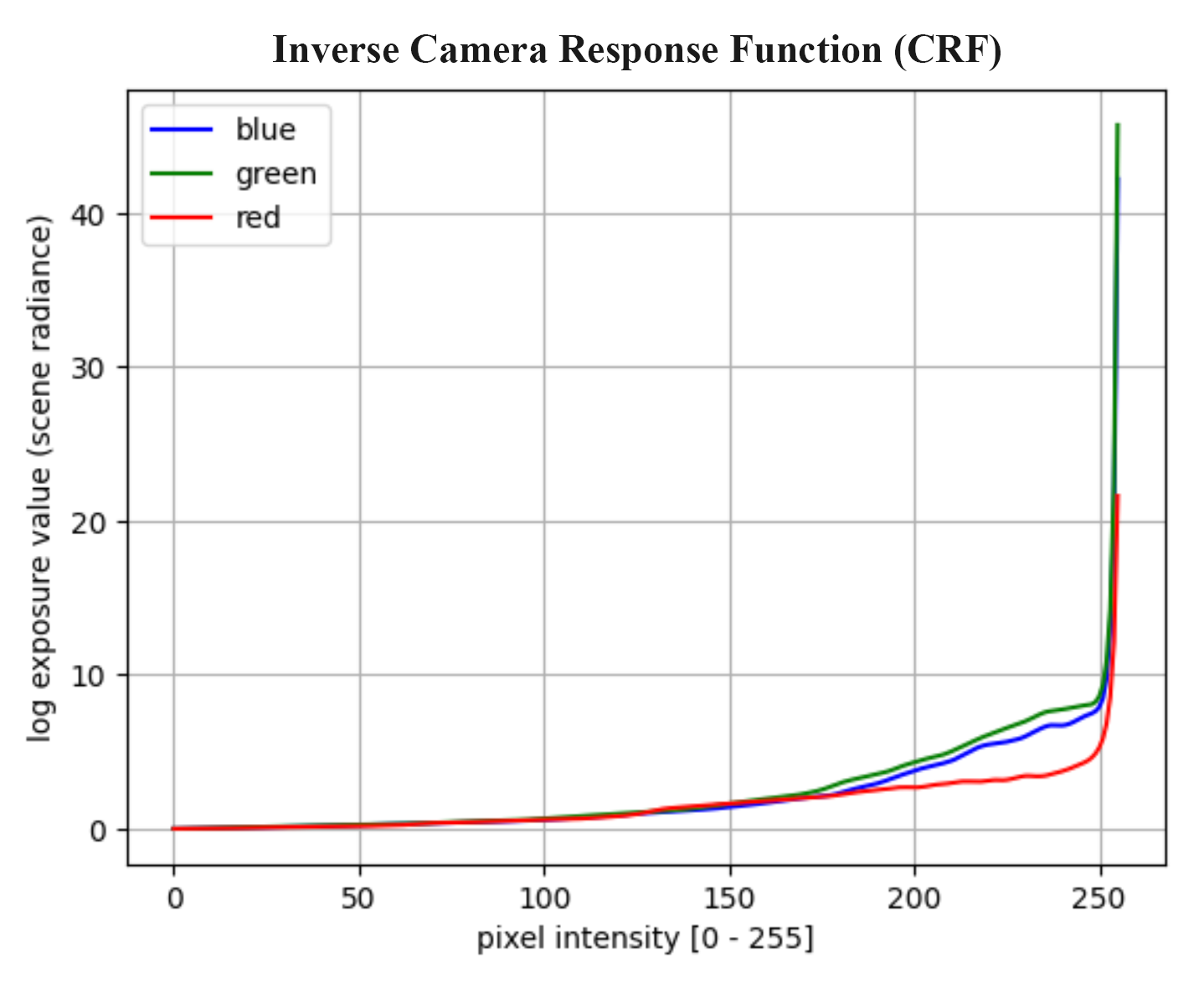}
\caption{\textbf{Inverse Camera Response Function (CRF)} showing the recovered nonlinear function mapping from sRGB 8-bit pixel color to linear 32-bit radiance.}
\label{fig:crf}
\end{figure}

\begin{figure}
\centering
\begin{tabular}{@{}c@{\hspace{1mm}}c@{\hspace{1mm}}c@{}}
    \includegraphics[width=0.3\linewidth]{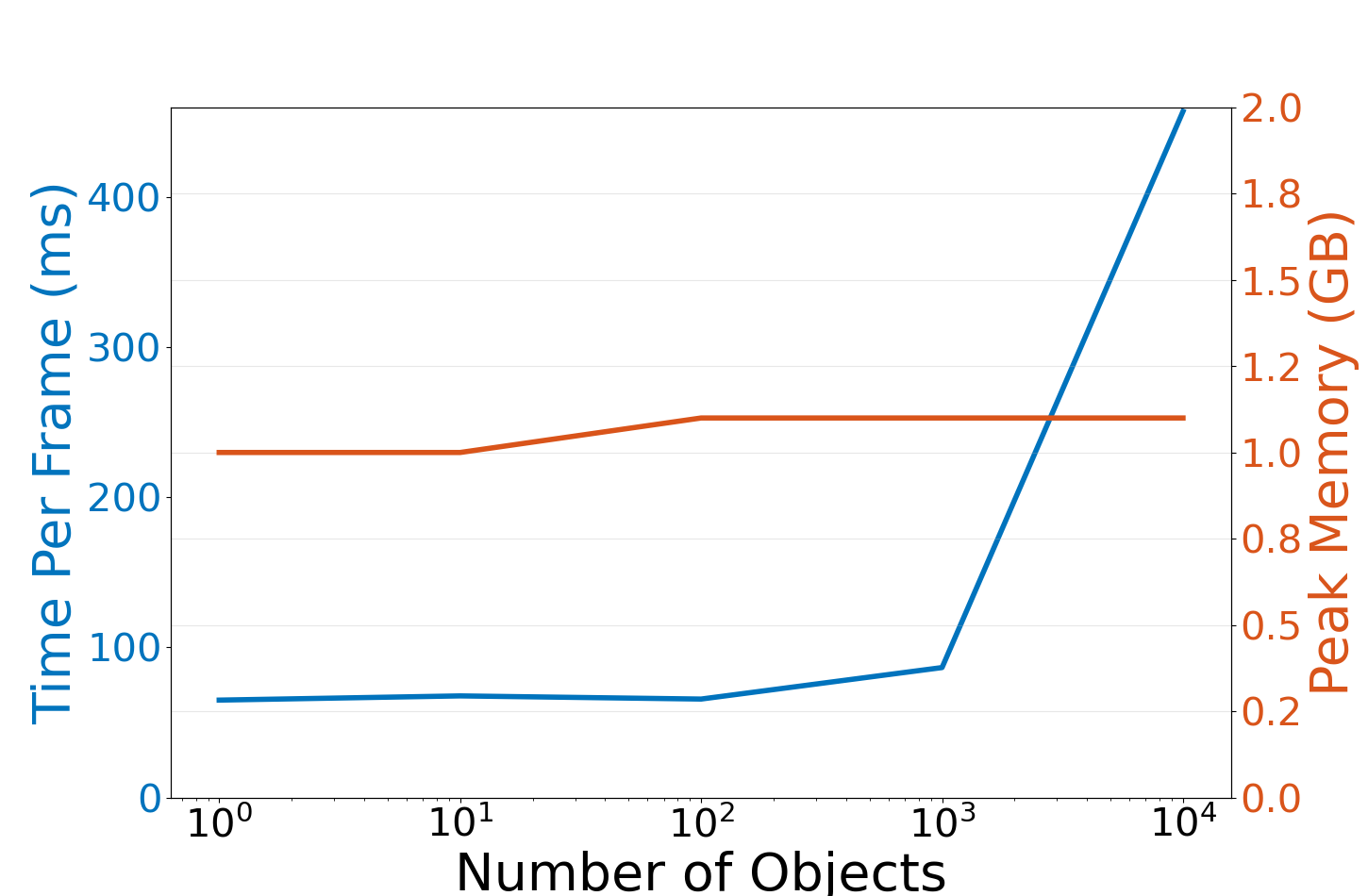} &
    \includegraphics[width=0.3\linewidth]{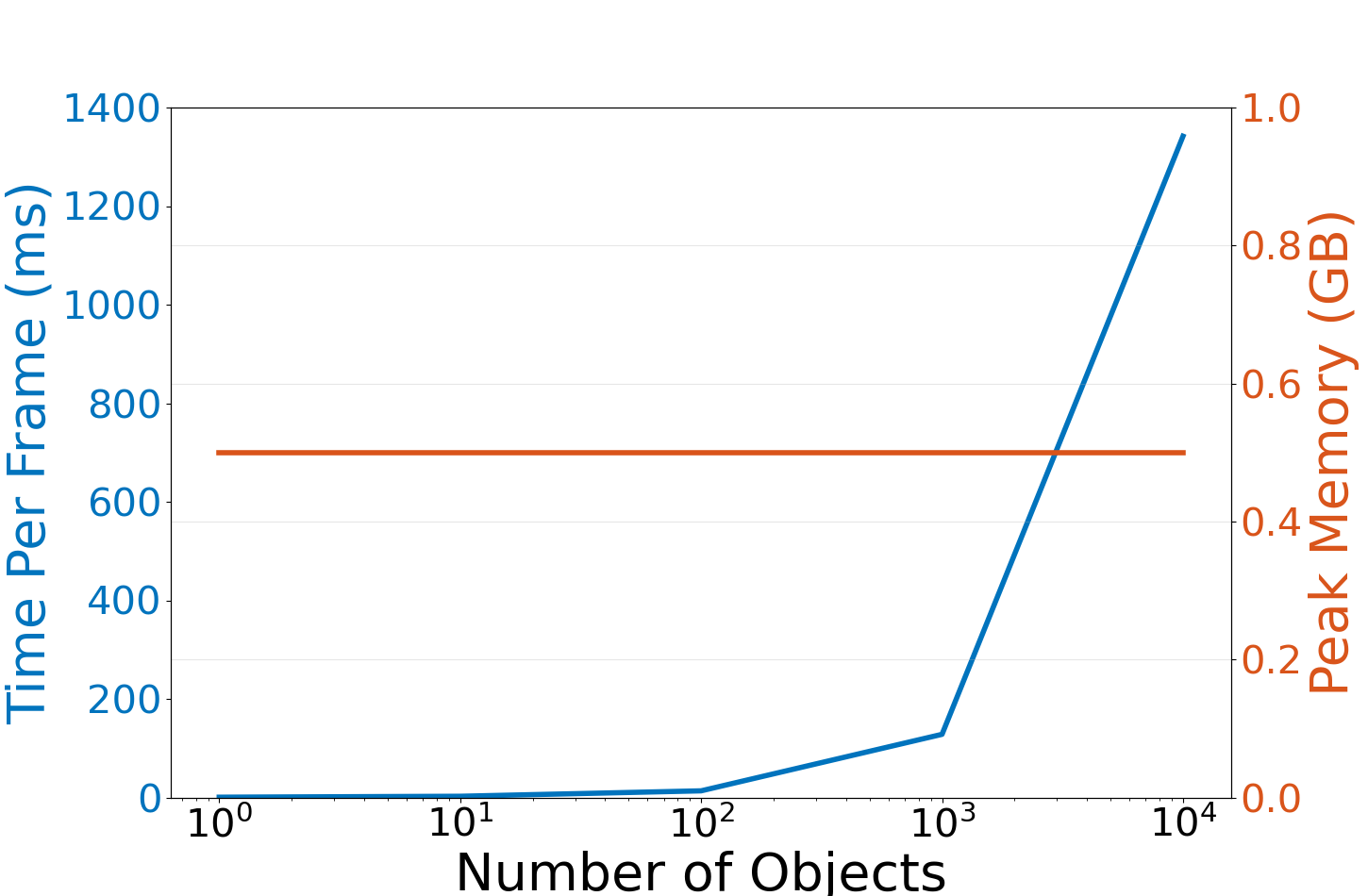}&
    \includegraphics[width=0.3\linewidth]{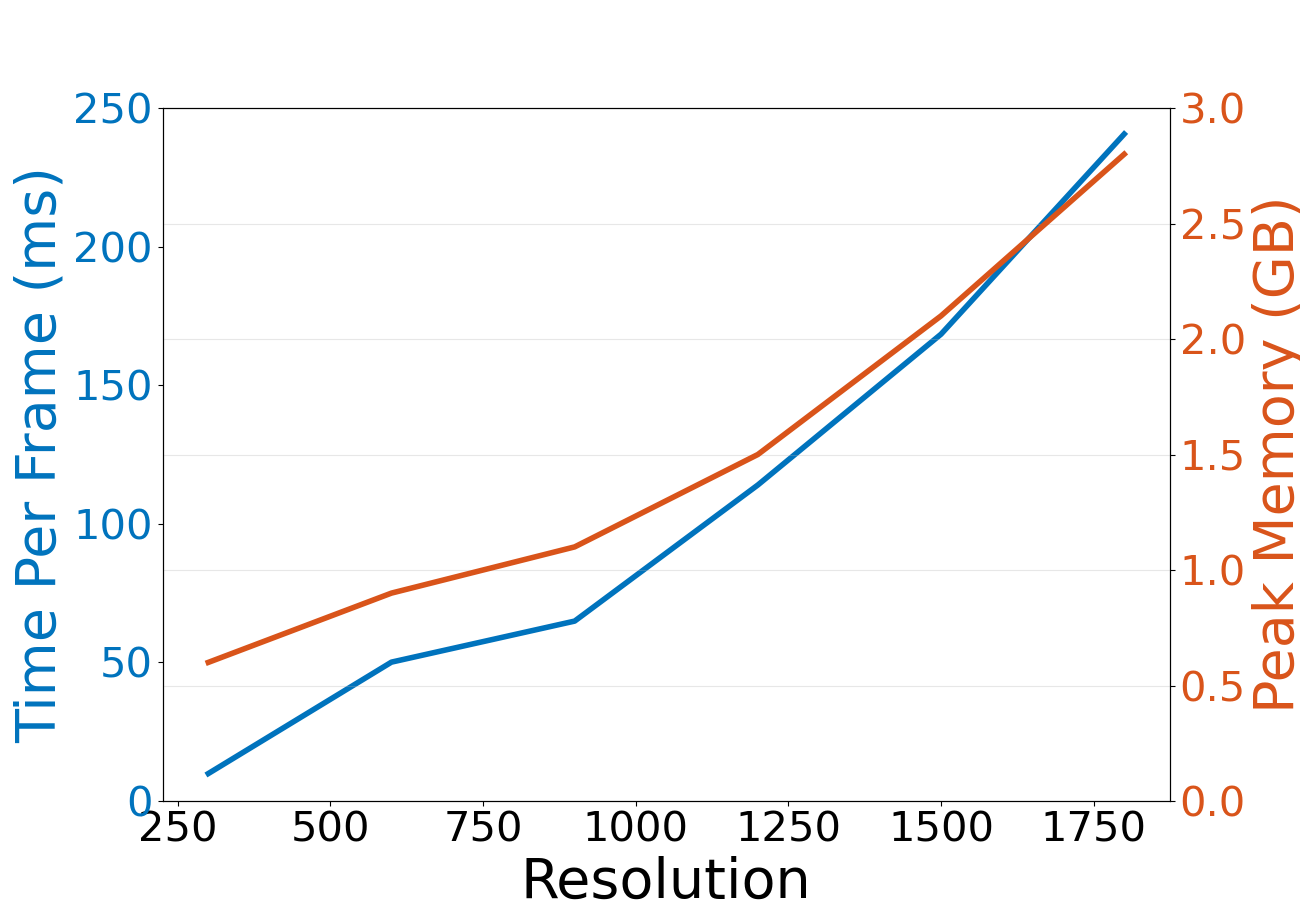}  \\
     \small (a) Rendering module  & \small (a) Simulation module & \small (c) Rendering module \\
\end{tabular}
\caption{\textbf{Scaling the number of the inserted objects and image resolution}. When we increase the number of simulated objects from $1$ to $10^4$, the peak simulation memory and rendering memory stay nearly constant. The (a) simulation time scales linearly w.r.t. the number of objects, while the (b) rendering time can increase when the number of the inserted meshes is too high. The (c) rendering time and memory usage in rendering modules scales linearly w.r.t. the number of pixels.
} 
\label{fig:number}
\end{figure}

\begin{figure}[ht]
 \includegraphics[width=0.3\linewidth]{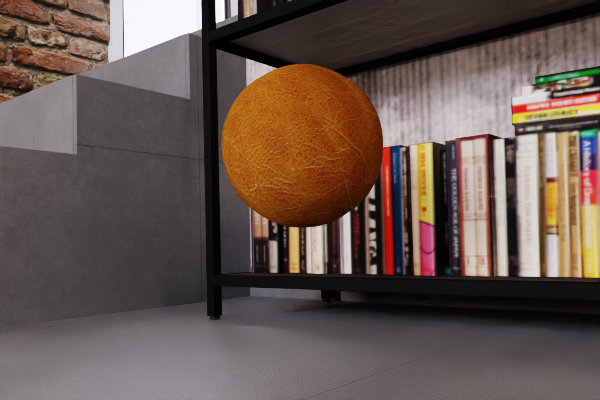}
\resizebox{0.4\linewidth}{!}{
\begin{tabular}[b]{l|cc}
	\toprule
	 & Rendering & Simulation  \\
	\hline
	Ours  & $\sim 0.1$ s & $\sim 0.02$ s  \\
	NeuPhysics & $\sim 5$ s  & $\sim 0.5$ s \\
	\bottomrule
\end{tabular}
}
% \vspace{-0.5em}
\caption{\textbf{Run time comparison with NeuPhysics~\cite{qiao2022neuphysics}}. We simulate a ball that bounces around (left). Compared to NeuPhysics, our method supports hybrid NeRF and mesh rendering and runs 25x-50x faster.}
\label{tab:speed_comparison}
% \vspace{-4mm}
\end{figure}

\begin{figure}
\centering
\begin{tabular}{@{}c@{\hspace{1mm}}c@{}}
    \includegraphics[width=0.3\linewidth]{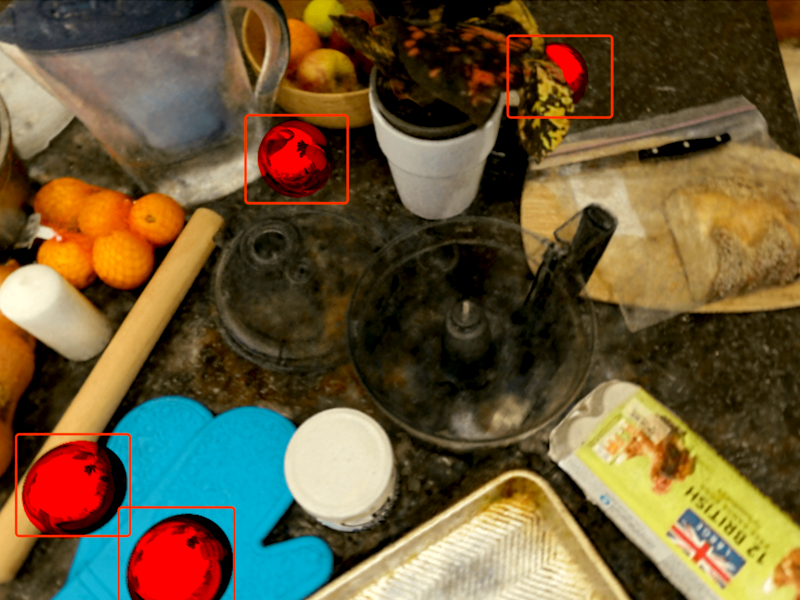} &
    \includegraphics[width=0.3\linewidth]{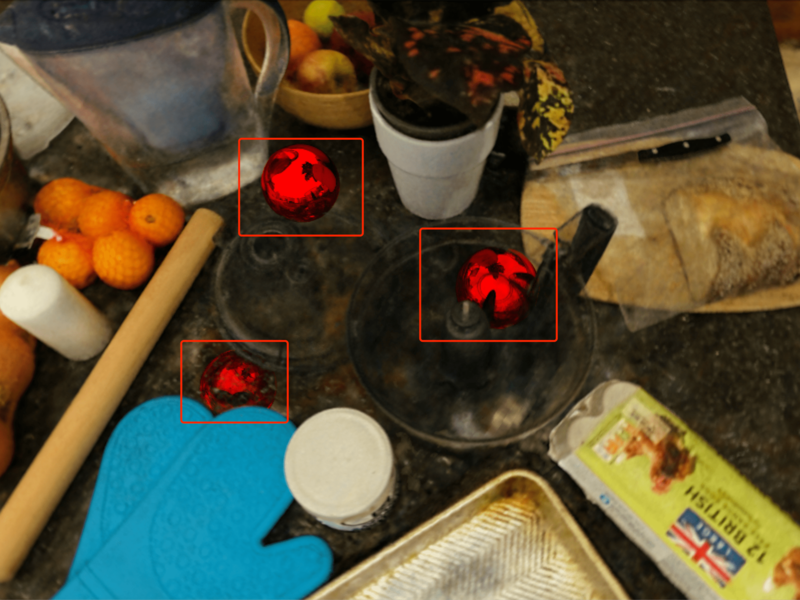}  \\
    \small (a) Using learned SDF to simulate & \small (b) Using mesh from density fields \\
    % \includegraphics[width=0.5\linewidth]{figure/geometry_ours.png} &
    % \includegraphics[width=0.5\linewidth]{figure/geometry_mc.png}  \\
    % \small (c) Our SDF Surface & \small (d) Marching-cube results
\end{tabular}
\vspace{-2mm}
\caption{\textbf{Simulation Comparison.} Our simulation (a) using learned SDF as collision proxy geometry can correctly handle the contact between the counter and four balls. (b) Using macrching-cube mesh directly from the density field, it fails to resolve collision and two of the four balls sink into the counter.}
\label{fig:comp_simulation}
\end{figure}

After capturing the scene, we preprocess the data using OpenCV.  For each bracketed exposure consisting of $n$ images taken from the same camera pose with different exposures, we align the images to ensure the best possible quality of the HDR image result, as even slight perturbations in the camera pose may cause degradation.  Then, we select a single bracketed series that is ``representative" of the lighting conditions of the full set, roughly covering the upper and lower limits of the overall dynamic range.  The representative bracketed series is used to recover the camera's nonlinear tone-mapping function, which we will refer to as the Camera Response Function (CRF) [Figure~\ref{fig:crf}] using the method from \cite{debevec1997}. 

To improve computational efficiency, we recover the CRF from only one representative image, although the linear system that must be solved could theoretically incorporate pixels from \textit{all} of the bracketed series as constraints. By applying the recovered CRF uniformly to all series, we achieve consistent levels across all images.  After applying the CRF to all bracketed series, we downsample each resulting 32-bit image by a factor of $4x$.  Finally, for numerical stability during downstream training, we normalize all of the images by linearly scaling radiance values to the range of $[0, 255]$.  While this may seem similar to reducing the HDR result back to LDR, all values are scaled uniformly so the \textit{relative} differences between values are preserved, and they are \textbf{not} quantized to integer values.  The resulting HDR images are then stored as EXR files.

Note that for performing camera pose estimation using COLMAP, as is widely done in the NeRF literature, we only use a single 8-bit LDR image from each bracketed series. 
 Similarly, when training an LDR model for comparison against its HDR counterpart, we use a single image from each bracketed series, where all of the singleton images have the same exposure duration.

\subsection{HDR NeRF Implementation Details}
Training the HDR NeRF is largely the same as training the baseline (LDR) NeRF.  In fact, the Instant-NGP implementation already supports training with 32-bit HDR images in EXR format.  We simply need to toggle flags `is\_hdr' in Instant-NGP's data loader and `linear\_colors' in the GUI.  Therefore, training is supervised purely in (scaled) linear color space, and correspondingly, the radiance component of the neural field is defined in linear color space.  Optionally, during rendering only, we may choose to apply a tonemapping function uniformly to all of the accumulated pixel radiance values to obtain the final image in sRGB color space.

\section{Details about the Rendering comparison}
\label{app:details_comparison}
% Here we show more details about the rendering comparison in Figure~\ref{fig:comp_render}. Figure~\ref{fig:nvdiffrec_womask} are the results when we want to use NVDiffRec to reconstruct the entire scene. The results are not good probably due to the complexity of the background. Therefore, we add a foreground mask and only optimize the foreground for comparison. These results might indicate that although neural fields are good at modeling background, it is difficult to export them as surface mesh. This is also our motivation to directly incorporate NeRF into the surface rendering pipeline.

\begin{figure}
\vspace*{-1.5em}
    \centering
    \includegraphics[width=0.5\linewidth]{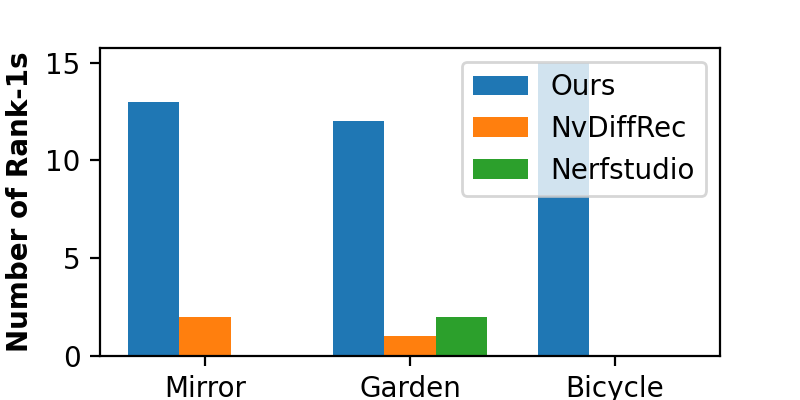}
    \vspace*{-1em}
    \caption{Number of Rank-1s each method receives.}
    \label{fig:user_study}
    % \vspace*{-1em}
\end{figure}

We conducted a user study where participants were asked to rank the realism, quality, and correctness of rendered results from three methods, including ours, on the following examples: Mirror (main paper Fig. 2), Garden, and Bicycle (main paper Fig. 5). 
Of 15 participants, 13/15 ranked our results highest for Mirror, while 12/15 did so for Garden and 15/15 for Bicycle (see Fig.~\ref{fig:user_study}). 
Binomial tests indicate that our images are statistically superior in all three scenes, with significant results at a significance level of 0.05. The corresponding p-values were found to be $0.035, 0.007$, and $6\times10^{-5}$, respectively.
We will include the quantitative results in the revised paper.

We also consider metrics such as FID and CLIP [5] scores. 
However, FID requires a ground truth distribution we cannot access. 
CLIP is a large language/vision model, but its scores are sensitive to prompts, and it is likely prone to misinterpreting reflections. 
Defining a scalable and robust quantitative metric to measure the visual fidelity of generative tasks remains an open problem.

MirrorNeRF designs a convenient capture system utilizing mirrors, but it does not focus on the rendering problem, thus does not overlap with the scope of our work.

\begin{figure}
\centering
\begin{tabular}{@{}c@{\hspace{1mm}}c@{}}
    \includegraphics[width=0.36\linewidth]{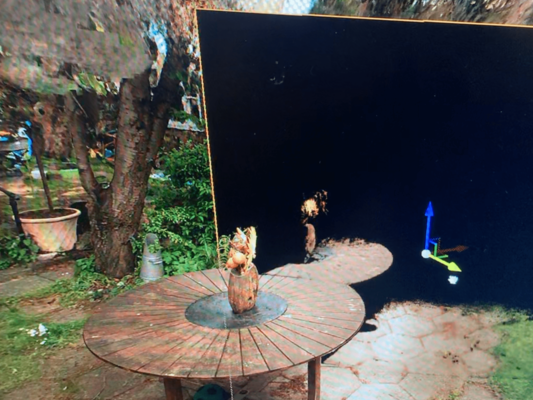}&
    \includegraphics[width=0.36\linewidth]{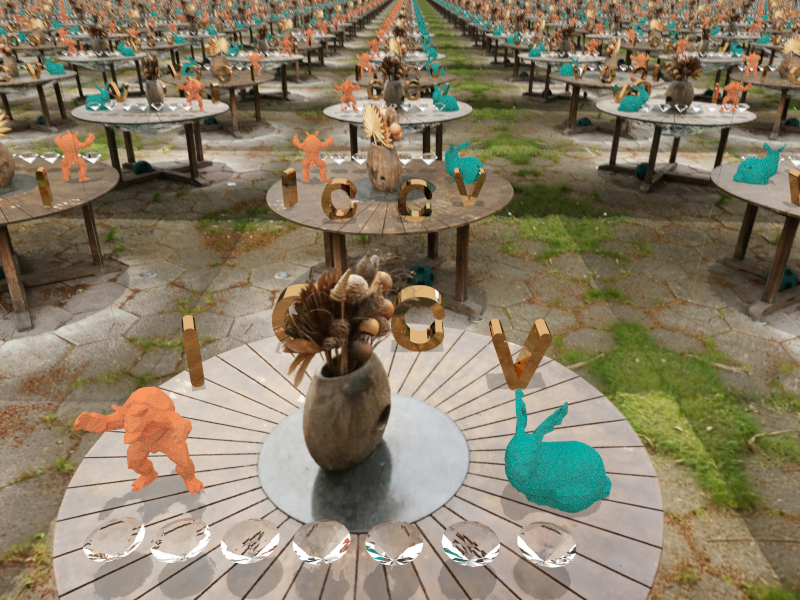} \\
    % \includegraphics[width=0.33\linewidth]{figure/driving2.jpg}  \\
    % \small (a) Luma AI & (b) Ours %& \small (c) Ours-shadow \\
\end{tabular}
\vspace{0mm}
\caption{Comparison between Luma AI (left) and Ours (right).}
\vspace{-2em}
\label{fig:luma}
\end{figure}

LumaAI recently released a \href{https://www.cgchannel.com/2023/04/use-nerfs-in-unreal-engine-with-luma-ais-new-plugin/}{UE plugin} in April \emph{after} the submission deadline. 
They provide binaries but no source code, so we only have a limited understanding of how it works. 
It is stated on their blog that NeRF cannot be used to simulate (otherwise, users need to add collider objects \emph{manually}). 
Moreover, when we put a mirror in the Garden, shadows in the mirror reflections look strange. 
By contrast, our reflections and shadows are superior (see Fig.~\ref{fig:luma}). 

\end{document}